\documentclass[3p,12pt]{elsarticle}

\usepackage{algorithm,algpseudocode}
\usepackage{amsmath,amsfonts,amsfonts,amsthm}
\usepackage{amssymb}
\usepackage[titletoc,title]{appendix}
\usepackage{bbm}
\usepackage[english]{babel}
\usepackage{blindtext}
\usepackage{caption}
\usepackage{courier}
\usepackage[usenames, dvipsnames]{color}
\usepackage{graphicx}
\usepackage[colorlinks]{hyperref}
\usepackage[nameinlink, capitalize]{cleveref}
\usepackage{lineno}
\usepackage{listings}
\usepackage[version=3]{mhchem}
\usepackage{siunitx}
\usepackage{subcaption}
\usepackage{todonotes}
\usepackage{url}
\biboptions{sort&compress}

\crefname{lstlisting}{Listing}{Listings}

\definecolor{airforceblue}{rgb}{0.36, 0.54, 0.66}
\definecolor{mygreen}{rgb}{0,0.6,0}
\definecolor{mygray}{rgb}{0.5,0.5,0.5}
\definecolor{mymauve}{rgb}{0.58,0,0.82}

\journal{Computer Physics Communications}

\theoremstyle{definition}

\numberwithin{equation}{section}

\def\p{\partial}
\newcommand{\f}[2]{{\frac{#1}{#2}}}
\newcommand{\wt}[1]{{\widetilde{#1}}}
\newcommand{\wh}[1]{{\widehat{#1}}}
\def\omegavec{{\mbox{\boldmath$\omega$}}}
\def\uvec{{\mbox{\boldmath$u$}}}
\def\phivec{{\mbox{\boldmath$\phi$}}}
\def\mvec{{\mbox{\boldmath$m$}}}

\graphicspath{{./Figures/}}
\begin{document}

\begin{frontmatter}
\title{A TensorFlow Simulation Framework for Scientific Computing of Fluid Flows on Tensor Processing Units}
\author[1]{Qing Wang\corref{cor1}}
\ead{wqing@google.com}
\author[1,2]{Matthias Ihme}
\author[1]{Yi-Fan Chen}
\author[1]{John Anderson}


\address[1]{Google, 1600 Amphitheatre Parkway, Mountain View, CA 94043, USA}
\address[2]{Stanford University, 440 Escondido Mall, Stanford, CA 94305, USA}

\cortext[cor1]{Corresponding author}

\begin{abstract}
A computational fluid dynamics (CFD) simulation framework for fluid-flow prediction is developed on the Tensor Processing Unit (TPU) platform. The TPU architecture is featured with accelerated dense matrix multiplication, large high bandwidth memory, and a fast inter-chip interconnect, making it attractive for high-performance scientific computing. The CFD framework solves the variable-density Navier-Stokes equation using a low-Mach approximation, and the governing equations are discretized by a finite-difference method on a collocated structured mesh. It uses the graph-based TensorFlow as the programming paradigm. The accuracy and performance of this framework is studied both numerically and analytically, specifically focusing on effects of TPU-native single precision floating point arithmetic. The algorithm and implementation are validated with canonical 2D and 3D Taylor-Green vortex simulations. To demonstrate the capability for simulating turbulent flows, simulations are conducted for two configurations, namely decaying homogeneous isotropic turbulence and a turbulent planar jet. Both simulations show good statistical agreement with reference solutions. The performance analysis shows a linear weak scaling and a superlinear strong scaling up to a full TPU v3 pod with 2048 cores.





\end{abstract}


\end{frontmatter}
\tableofcontents
\section{\label{SEC_INTRO}Introduction}
Computational fluid dynamics (CFD) has become an indispensable tool in the scientific community and in industry. CFD simulations are now widely employed for analyzing turbulent flows, for predicting heat transfer and combustion processes in complex geometries, in augmenting expensive hardware tests, and in studying environmental flows, among others~\cite{NASA_VISION2030}. These capabilities have been enabled through the remarkable progress in the development of computational algorithms, the construction of physical models, and advances in data analytics. In addition, dramatic improvements in the availability, speed, and efficiency of high-performance computing (HPC) systems have enabled high-fidelity flow simulations with increasing resolution and considering complex physical processes.

Algorithms and discretization schemes employed for flow simulations have been developed to map well onto traditional computing systems with structured hardware architectures. However, the large data transfer and memory access results in relatively low computational intensity, and algorithmic advances are necessary to take advantage of specialized and heterogeneous computing architectures~\cite{DONGARRA_LASTOVETSKY_BOOK2009,Brodtkorb_Dyken_Hagen_Hjelmervik_Storaasli_SP2010,TERZO_DJEMAME_SCIONTI_PEZUELA_BOOK2020}. In particular, graphics processing units (GPUs), originally developed for accelerating graphics and imaging processing, have gained significant interest for high-performance scientific computing. Substantial performance gains over multi-core CPU computations have been demonstrated on GPU hardware due to their high memory bandwidth and computational throughput~\cite{HARRIS_GPUGEMS2004,BRANDVIK_PULLAN_PIME2007,ELSEN_LEGRESLEY_DARVE_JCP2008,RAVIKUMAR_APPELHANS_YEUNG_PICHPCNSA2019}. 

Increasing interest in artificial intelligence has stimulated the development of the tensor processing unit (TPU) as an  application-specific integrated circuit (ASIC) for accelerating machine learning (ML) applications~\cite{JOUPPI_ETAL_ACM2017}. While TPUs have been primarily employed for deep learning, their highly specialized architecture is also attractive for scientific computing. Specifically, the design of the TPU architecture for performing highly efficient dense matrix multiplications makes it suitable for flow simulations, involving the solution of pressure Poisson systems and high-order discretizations. The high-bandwidth interconnect and large memory hold promise for reducing latency and accommodating storage requirements to deal with problems of higher computational intensity. Furthermore, the abstraction of the TPU architecture behind the TensorFlow framework~\cite{TENSORFLOW_2015} provides direct access to software libraries and application programming interfaces (APIs) for scientific computing and ML applications. 

While the utilization of TPUs for HPC has been demonstrated in applications to discrete-event simulations~\cite{BELLETTI_ETAL_ARXIV2019,YANG_CHEN_ROUMPOS_COLBY_ANDERSON_ARXIV2019} and Fourier transformation~\cite{LU_CHEN_HECHTMAN_WANG_ANDERSON_ARXIV2020}, their extension to solving partial differential equations, such as those that describe complex flows, has not been explored. This can largely be attributed to the overhead associated with refactoring existing CFD algorithms to map well on these graph-based computing architectures, and specialized floating-point arithmetic that was specifically developed for ML applications. The objective of this work is to evaluate the application of TPUs in conjunction with the TensorFlow programming environment for scientific computing of fluid flows with the specific goal of targeting high-fidelity flow simulations. To this end, we implement a CFD algorithm on TPU with TensorFlow for low-Mach number flows with variable density. We evaluate its performance with focuses on the accuracy and scalability.

The remainder of this article has the following structure. \Cref{SEC_TPU} introduces the TPU architecture and its application. The governing equations, numerical algorithms and implementation details specific to the TPU architecture are presented in ~\cref{SEC_MATH_MODEL}, while additional considerations regarding the floating-point precision are presented in~\cref{SEC_SINGLE_PRECISION_ARITH}. The resulting algorithm is evaluated and its performance is assessed in~\cref{SEC_RESULTS,SEC_PERFORMANCE}. For this, four configurations of increasing complexity are considered, which include Taylor-Green vortex flow in two and three dimensions, decaying homogeneous isotropic turbulence, and a turbulent planar jet flow. The article finishes by summarizing the main findings and offering conclusions in~\cref{SEC_CONCL}.

\section{\label{SEC_TPU}Tensor processing unit}
\begin{figure}[!htb!]
  \centering
  \begin{subfigure}[b]{0.42\textwidth}
    \includegraphics[width=\textwidth]{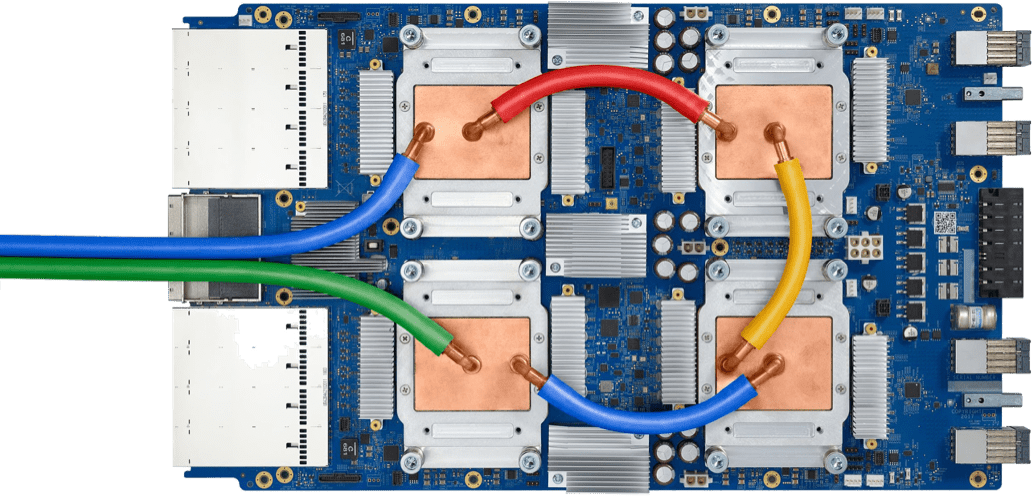}
    \caption{\label{FIG_TPU_BOARD}TPU v3 with four chips.}
  \end{subfigure}
  \hfill
  \begin{subfigure}[b]{0.57\textwidth}
    \includegraphics[width=\textwidth]{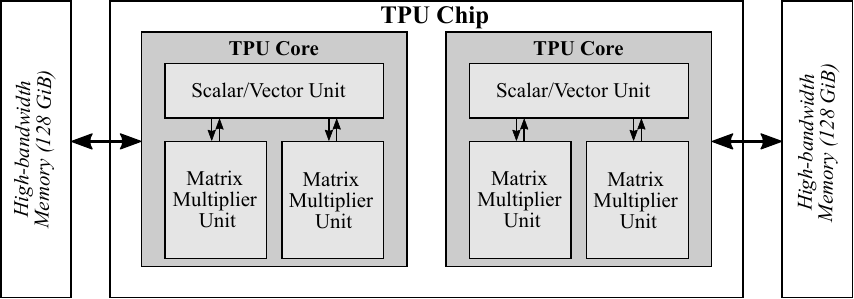}
    \caption{\label{FIG_TPU_CHIP}Diagram of TPU-chip architecture.}
  \end{subfigure}
  \caption{\label{FIG_TPU_BOARD_CHIP}Tensor Processing Unit (TPU v3): (a) TPU board with four cores, (b) diagram of chip architecture, consisting of two cores, each chip contains two cores, and each core is equipped with a scalar processor, a vector processor, and two matrix multiplier units; arrows indicate connections across processing units and memory access.}
\end{figure}

The basic unit of the TPU v3 architecture, which is considered in this work, consists of a TPU board with four independent chips that are associated with one CPU host, see~\cref{FIG_TPU_BOARD_CHIP}. Each chip has two tensor compute cores that are optimized for vectorized operations and dense-matrix operations. Up to 1024 chips are connected through a dedicated high-speed, low-latency, two-dimensional toroidal inter core interconnect (ICI) network, forming a TPU pod.~\Cref{tab:tpu_features} shows key features of the TPU v3 architecture~\citep{Jouppi2021-hf}.

\begin{table}[!hb!]
\centering
\footnotesize
\caption{Key features of the TPU v3 architecture.}
\begin{tabular}{|c|c|}\hline
    Peak TFLOPS/Chip & 123 (bfloat16) \\
    \hline
    Network links$\times$Gbits/s/Chip & $4\times 656$  \\
    \hline
    Max Chips/Pod & 1024 \\
    \hline
    Cores/Chip & 2 \\
    \hline
    Chips/CPU Host & 4 \\
    \hline
    Memory Size (On-/Off-Chip) & 32MB/32GB \\
    \hline
    Memory Bandwidth GB/s/Chip & 900 \\
    \hline
    MXU Size/Core & 2 ($128\times 128$) \\
    \hline
 \end{tabular}
 \label{tab:tpu_features}
\end{table}

Each TPU core has 16 GiB on-chip high-bandwidth memory (HBM) and consists of a scalar, a vector, and two matrix multiplication units (MXU). The HBM is associated with a specific TPU core, and is not shared across cores and chips. Each MXU supports $128\times 128$ multiply-accumulate (MAC) operations per cycle, with a raw peak throughput of 22.5 teraflops. While the inputs and outputs of the MXU are single precision (float32), MAC operations are typically performed with a bfloat16 (16-bit) floating point representation. The bfloat16 format uses a 8-bit exponent and 7-bit mantissa to provide a larger range than the IEEE half-precision (float16) representation for deep learning. Software emulation with multiple passes of the MXU operations makes higher-precision matrix multiplication (32-bit and 64-bit) possible, with a performance trade-off~\cite{KALAMKAR_ETAL_ARXIV2019}. In contrast, the vector and scalar hardware units support native operators at float32, but software emulation is needed for float64. The results reported in this work are all performed with float32, and the impact of single-precision arithmetic will be examined.

The TPU board is connected through the PCIe-bus to the host server, which sends the instruction to the TPU for execution. The ICI network directly connects the cores within a TPU pod, so that the communication between TPU cores goes directly through this dedicated network without involving the PCIe-bus or host.

The TPU software stack is programmed either through TensorFlow~\cite{TENSORFLOW_2015}, JAX~\cite{jax2018github}, or PyTorch~\cite{PYTORCH_2019}. In the case of a TensorFlow  implementation, which is considered in the present work, a computational graph is compiled just-in-time in the beginning of the run. The graph is compiled and partitioned on the host CPU, and is sent to the TPU for execution. The compilation and optimization of the TPU-executable subgraph is done by the Accelerated Linear Algebra (XLA) compiler, which handles the generation of the TPU-specific code for data management, on-chip memory allocation, hardware execution on the scalar, vector and MXU units, and inter-chip communication~\cite{JOUPPI_ETAL_ACM2017}. 

\section{\label{SEC_MATH_MODEL}Mathematical model and implementation}
This work is concerned with solving the variable-density Navier-Stokes equations on TPU architectures. In the following, an overview of the governing equations and the discretization is provided. This is followed by discussing salient aspects of the TensorFlow TPU implementation.
\subsection{\label{SSEC_GOV_EQUATIONS}Governing equations}
The continuity and momentum equations that describe the fluid motion are written as: 
\begin{subequations}
 \label{EQ_GOVERNING_ALL}
 \begin{align}
 \label{EQ_GOVERNING_MASS}
 \p_t \rho+ \nabla\cdot(\rho \boldsymbol{u}) &= 0\;,\\
 \label{EQ_GOVERNING_MOM}
 \p_t(\rho \boldsymbol{u}) +\nabla\cdot(\rho\boldsymbol{u}\otimes\boldsymbol{u}) &= -\nabla p + \nabla\cdot\boldsymbol{\sigma}\;,
    \end{align}
\end{subequations}
where $\rho$ is the density, $\boldsymbol{u}$ is the velocity vector, $p$ is the pressure, and $\boldsymbol{\sigma}$ is the stress tensor,
\begin{equation}
\boldsymbol{\sigma} = 2\mu\boldsymbol{S}-\f{2}{3}\mu(\nabla\cdot\boldsymbol{u})\bf{I}\;,
\end{equation}
where $\mu$ is the dynamic viscosity, $\bf{I}$ is the identity matrix, and the strain rate is:
\begin{equation}
\boldsymbol{S}=\frac{1}{2}\left[\nabla\boldsymbol{u}+(\nabla\boldsymbol{u})^T\right].
\end{equation}
To consider scalar mixing,~\cref{EQ_GOVERNING_ALL} is augmented by solving a scalar transport equation, which takes the following form:
\begin{equation}
\label{EQ_GOVERNING_SCALAR}
     \p_t(\rho \phi) + \nabla\cdot(\rho \boldsymbol{u}\phi) =  \nabla\cdot(\rho\alpha\nabla\phi)\;,
\end{equation}
with $\alpha$ being the scalar diffusivity and $\phi\in[0,1]$ is a conserved scalar. In the present work, a scalar mixing law is considered to relate the  density to the scalar, which is denoted as function $f$:
\begin{equation}
 \label{EQ_EOS}
 \rho = f(\phi)\;.
\end{equation}
\subsection{\label{SSEC_NUM_ALG}Numerical algorithm}
The set of governing equations is solved on a Cartesian coordinate system employing an equidistant mesh in each directions. All dependent variables are stored on a collocated mesh, and spatial derivatives are discretized using a finite-difference scheme. Time-staggering is employed for the temporal discretization. For this, transported scalars and the density are stored at full time steps, $t=t^{n}$, and velocity and pressure are stored at half time steps $t={t^{n-1/2}}$ that are behind the scalar and density. The solution is advanced in time using a fractional step  method~\cite{KIM_MOIN_JCP1985}, having second-order accuracy in time.

The algorithm employs a time-explicit iterative scheme~\cite{Pierce2001-nw}. In the following, the subscript $l$ refers to the subiteration cycle, which provides an estimate to the value at the updated time-step $t^{n+1}$. A provisional value is denoted by $\wh{\cdot}$. Prior to each subiteration $l$, all boundary conditions for the state vector and pressure are updated. The solution at $l=0$ is initialized with the available solution at the current time step, that is: 
\begin{equation}
   \boldsymbol{u}^{n+1/2}_{l=0} = \boldsymbol{u}^{n-1/2}\;,\quad
   \phi^{n+1}_{l=0} = \phi^n\;,\qquad
   \rho^{n+1}_{l=0} = \rho^n\;.
\end{equation}
\paragraph{Scalar and density update} To account for variations in density that are considered through the state relation~\cref{EQ_EOS}, we first advance~\cref{EQ_GOVERNING_SCALAR} to solve
for the scalar density $\Phi=(\rho\phi)$ at time $t^{n+1}$. Using a  semi-discrete formulation in which we approximate the temporal derivative by a central-difference scheme, this can be written as:
\begin{equation}
    \frac{\Phi^{n+1}_{l+1}-\Phi^n}{\Delta t} = -  \nabla\cdot\left[\left\{\rho\right\}^{n+1/2}_{l} \boldsymbol{u}_l^{n+1/2}\left\{\phi\right\}^{n+1/2}_{l}\right] + \nabla\cdot \left[\left\{\rho\right\}^{n+1/2}_{l}\left\{\alpha\right\}_{l}^{n+1/2}\nabla\left\{\phi\right\}^{n+1/2}_{l}\right]\;,
    \label{EQ_SCALAR_UPDATE}
\end{equation}
where we introduced the operator $\left\{\psi\right\}_l^n=\frac{1}{2}(\psi_l^{n+1/2}+\psi^{n-1/2})$ to denote temporal interpolation of time-staggered data. \Cref{EQ_SCALAR_UPDATE} provides an updated estimate for $\Phi_{l+1}^{n+1}$. This allows us to evaluate a provisional value for the scalar $\wh{\phi}={\Phi^{n+1}_{l+1}}/{\rho_l^{n+1}}$, from which the density and the scalar are computed:
\begin{equation}
 \label{EQ_RHO_UPDATE}
 \rho_{l+1}^{n+1} = f(\wh{\phi})\qquad\text{and}\qquad
 \phi_{l+1}^{n+1} = {\Phi^{n+1}_{l+1}}/{\rho^{n+1}_{l+1}}
\end{equation}
\paragraph{Momentum update} With the updated density and scalar from~\cref{EQ_RHO_UPDATE}, a provisional estimate for the momentum flux $\wh{\mvec}_{l+1}$ at $t^{n+1/2}$ is obtained by time-advancing~\cref{EQ_GOVERNING_MOM}:
\begin{equation}
    \label{EQ_MOM_UPDATE}
   \frac{\wh{\mvec}_{l+1}-\mvec^{n-1/2}}{\Delta t} = -  \nabla\cdot\left[\left\{\mvec\right\}^{n}_{l} \left\{\uvec\right\}^{n}_{l}\right] -\nabla p^n + \nabla\cdot \left[\left\{\boldsymbol{\sigma}\right\}_{l}^{n}\right]\;,
\end{equation}
with $\mvec=\rho\uvec$. From this, an intermediate solution for the velocity is computed: 
\begin{equation}
 \label{EQ_VEL_PREDICTOR}
 \wh{\uvec}_{l+1} = \frac{\wh{\mvec}_{l+1}}{\left\{\rho\right\}^{n+1/2}_{l+1}}\;. 
\end{equation}

\paragraph{Pressure correction}
Continuity is enforced by solving a Poisson equation in which the pressure is corrected to fulfill mass conservation. This is achieved by writing the momentum-flux and pressure updates as:
\begin{subequations}
 \label{EQ_PPROJ_UPDATES}
 \begin{align}
 \label{EQ_PPROJ_UPDATES_MOM}
 \mvec^{n+1/2}_{l+1} &=\wh{\mvec}_{l+1} + \delta \mvec\;,\\
 \label{EQ_PPROJ_UPDATES_PRESS}
 p^{n}_{l+1} &=p^n + \delta p\;.
    \end{align}
\end{subequations}
Upon inserting~\cref{EQ_PPROJ_UPDATES} into the semi-discrete momentum equation and subtracting \cref{EQ_MOM_UPDATE} gives
\begin{equation}
 \label{EQ_DELTA_UPDATE}
    \frac{\delta \mvec}{\Delta t} = - \nabla \delta p\;.
\end{equation}
By taking the divergence of~\cref{EQ_DELTA_UPDATE} and using the semi-discrete form of~\cref{EQ_GOVERNING_MASS} to evaluate $\nabla\cdot\mvec^{n+1/2}_{l+1}$ it follows
\begin{equation}
 \label{EQ_PRESS_POISSON_EQUATION}
    \nabla^2\delta p = \f{1}{\Delta t}\left[\f{\rho^{n+1}_{l+1}-\rho^n}{\Delta t} + \nabla\cdot \wh{\mvec}_{l+1}\right]\;.
\end{equation}
With $\delta p$ computed from the previous step, the velocity in~\cref{EQ_VEL_PREDICTOR} is then updated as:
\begin{equation}
 \uvec^{n+1/2}_{l+1}=\frac{\mvec_{l+1}^{n+1/2}}{\left\{\rho\right\}_{l+1}^{n+1/2}}\;,
\end{equation}
where the momentum flux is determined from
\begin{equation}
 \mvec^{n+1/2}_{l+1} = \wh{\mvec}_{l+1}-\Delta t\nabla\delta p\;.    
\end{equation}
\paragraph{Update boundary conditions}
The last step in the sub-iteration loop is to update the boundary conditions for velocity and scalars. The pressure boundary conditions are enforced at the beginning of each sub-iteration.
\subsection{\label{SSEC_NUM_DISCRETIZATION}Remarks on discretization}
\paragraph{Spatial discretization}
In the present work, all spatial operators are discretized using a finite-difference formulation on a collocated mesh. The advection term in the scalar transport equation is discretized using the QUICK scheme~\cite{Leonard1979-cd}. 

To eliminate the checker-board effect, arising from evaluating the fluxes at the cell faces, a Rhie-Chow correction~\cite{RHIE_CHOW_AIAAJ1983} is employed. Using $ijk$-index notation to enumerate the grid points in the spatial discretization (and omitting explicit notation for the time step discretization), the momentum at the cell face $x_{i+1/2}$ takes the form:
\begin{equation}
    (m_j)_{i+1/2} = \frac{1}{2}\left[(m_j)_i + (m_j)_{i+1}\right] - \frac{\Delta t}{4\Delta x}\left(-p_{i-2} + 3p_{i-1} - 3p_{i} + p_{i+1}\right).
\end{equation}
where $m_j=\rho u_j$.

Using the QUICK scheme, the scalar flux at $x_{i+1/2}$ is then evaluated as:
\begin{equation}
    (m_j \phi)_{i+1/2} = \frac{1}{2} \left[(m_j)_{i+1/2} + \left|(m_j)_{i+1/2}\right|\right]\phi^+ + 
    \frac{1}{2} \left[(m_j)_{i+1/2} - \left|(m_j)_{i+1/2}\right|\right]\phi^-\;,
    \label{eq:quick_flux}
\end{equation}
where:
\begin{subequations}
\begin{align}
    \phi^+ &= \frac{1}{8}\left[-\phi_{i-1} + 6\phi_i + 3\phi_{i+1}\right]\;,\\
    \phi^- &= \frac{1}{8}\left[3\phi_i + 6\phi_{i+1} - \phi_{i+2}\right]\;.
\end{align}
\end{subequations}
The convection terms and the divergence of the momentum are then computed as:
\begin{subequations}
\begin{align}
  \partial_{i}(m_j \phi) &= \frac{(m_j \phi)_{i+1/2} - (m_j \phi)_{i-1/2}}{\Delta x}\;,\\
  \partial_{i}(m_i) &= \frac{(m_i)_{i+1/2} - (m_i)_{i-1/2}}{\Delta x}\;.
\end{align}
\end{subequations}
where $\partial_{i}=\partial/\partial x_i.$
\paragraph{Temporal discretization}
An iterative approach is employed for the time advancement. As discussed in~\cref{SSEC_NUM_ALG}, this is achieved by performing sub-iterations for each time step. In each sub-iteration, a prediction for variables at the next time step $t^{n+1}$ is made. Together with the solution at the current time step $t^n$, the mid-point value at $t^{n+\frac{1}{2}}$ is computed. Similarly, interpolation is employed to evaluate the right-hand side of the momentum update at the midpoint condition $t^n$, see~\cref{EQ_MOM_UPDATE}. We proceed this iteration until desired convergence is reached. Note that we adopted this iterative approach instead of the classical Crank-Nicolson scheme that is fully implicit. The benefit of this approach is that we can perform the time integration of the transported scalars and velocity in a time-staggered manner, which was shown to have better convergence properties~\cite{Pierce2001-nw}.

\subsection{\label{SSEC_TPU_IMPL}TPU implementation}
\begin{figure}[!htb!]
  \centering
  \includegraphics[width=0.85\textwidth]{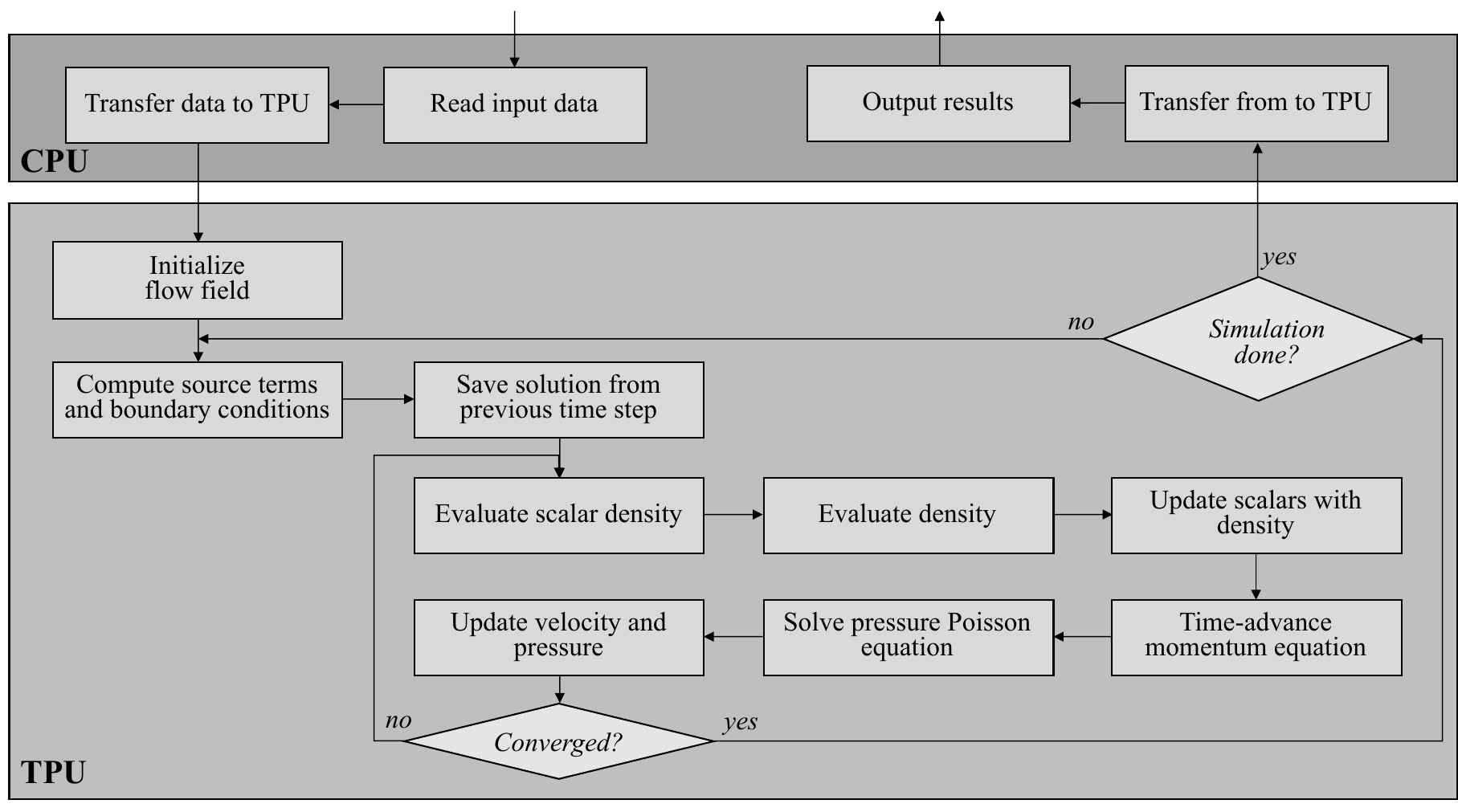}
  \caption{Flowchart of TPU solver implementation.}
  \label{fig:solver_algo}
\end{figure}

\Cref{fig:solver_algo} shows the flowchart of the algorithm implementation. All computations are done on the TPU. In the following, we present details on the parallelization, data structure, and graph-specific TensorFlow implementation.
\subsubsection{Parallelization}
The single-instruction, multiple-data approach is adopted for parallelization of the TPU solver implementation. The distribution of tasks to multiple cores is achieved through the \lstinline{tpu.replicate} API in the TensorFlow library. The partitioned data is stored in the HBM associated with each TPU core. Data communication between TPU cores on different chips is performed through the ICI at a data rate of 656 Gb/s~\citep{Jouppi2021-hf}.

In the current implementation, the computational domain is partitioned along the three directions using $P_x\times P_y\times P_z$ TPU-cores. Each Cartesian partition with $\wh{N}_x=N_x/P_x, \wh{N}_y=N_y/P_y$, and $\wh{N}_z=N_z/P_z$ is represented as a so-called TPU replica. Each TPU core has a unique replica, which maps to a specific location in the computational grid following the order of $x$, $y$, and $z$. The global index of each mesh point is then a linear function of the coordinates of the TPU replica in the computational grid. For example, the global index of a location in the TPU replica $(P_i, P_j, P_k)$ in the $x$-direction is computed as $i=\wh{N}_x (P_i-1) + \wh{i}$, where $\wh{\cdot}$ denotes a quantity that is local to a TPU-replica. 

The communication of boundary conditions between replicas is achieved through the TensorFlow operator \lstinline{collective_permute}. This operation is performed twice along each spatial direction by separating the communication along positive and negative directions. In each direction, a TPU-core passes a slice of a tensor-data field to its neighbor, and receives a slice of the same size. The data structure, discussed next, is laid out to optimize the parallelization along the $z$-direction and taking advantage of the TensorFlow intrinsic convolution through batch evaluations along the $x$- and $y$-directions. 
\subsubsection{\label{sssec:kernel_op}Data structures and kernel operations}
The data structure of the solver is laid out to achieve optimal MXU utilization and optimized graph parallelization on TPU architectures. To this end, we represent all three-dimensional state variables as python lists of 2D tensors \lstinline{tf.Tensor}, which are denoted as $[\phivec_{ij}]_k$ with indices $i,j$ and $k$ corresponding to the respective direction along the $x$-, $y$-, and $z$-coordinates. Considering the data structure after parallelization on a single TPU core, the outer dimension of this python list, which is indexed by $k$, is $\dim \{[\phivec_{ij}]_k\} = \wh{N}_z$, and the dimension of the 2D tensor \lstinline{tf.Tensor} is $\dim \{\phivec_{ij}\} = \wh{N}_x\times \wh{N}_y$, representing the $x$-$y$ plane. Note that for clarity, the explicit declaration of ghost cells and boundary points is omitted; instead, they are augmented during the run-time initialization. It's worth noting that this data structure provides better performance than the representation as a 3D \lstinline{tf.Tensor} for the TPU HBM architecture. On the one hand, the XLA compiler was designed to optimize operations for 2D-tensor structures~\citep{Jouppi2020-te}. On the other hand, depending on the associated operations for a given 3D \lstinline{tf.Tensor}, the XLA compiler rearranges the internal order of its axes and changes its underlying layout. This rearrangement leads to an inefficient padding in memory, which degrades performance. The list of 2D \lstinline{tf.Tensor} representation enforces the desired layout with more predictable memory and performance efficiency. However, with the continuing improvements of the XLA compiler, we anticipate that the 3D \lstinline{tf.Tensor} implementation will become more efficient and further help to simplify the data representation.

Provided that 3D tensors are represented as lists of 2D \lstinline{tf.Tensor}, we use directional splitting and implement discretized operators differently along directions of the \lstinline{tf.Tensor} field (the $x$- and $y$-axes) and the list coordinate (the $z$-axis). These implementations of discrete operations along the $x$-, $y$-, and $z$-axes, however, are mathematically identical, so that there is no operator-bias with respect to the spatial direction.

In the $x$- and $y$-directions, we map our algorithm to the TensorFlow API and represent spatial-derivative operators as one-dimensional convolutions. This is efficiently evaluated using TensorFlow's function \lstinline{tf.nn.conv1d}, which operates on an input tensor field of rank 3 with dimensions of batch size $M_\text{b}$ (\lstinline{batch}), input channels $M_{c}$ (\lstinline{in_channels}), and input width $M_\text{w}$ (\lstinline{in_width}). The weights of finite-difference operators are customized as filters in this function. Specifically, given a 2D tensor $\phivec_{ij}\in \mathbb{R}^{\wh{N}_x\times \wh{N}_y}$, we first transpose it to $\phi_{i'j'}\in \mathbb{R}^{M_\text{b}\times M_\text{d}}$, where $M_\text{d} = M_\text{c} \times M_\text{w}$ is the data size of a batch. Note that in $\phi_{i'j'}$, $j'=i$ if the operator is evaluated along the $x$-direction and $j'=j$ for the $y$-direction. This allows us to represent the first dimension as a ``batch'' in \lstinline{tf.nn.conv1d}, and repeated operations are applied along this dimension.

Denoting $\phivec_{i',:}$ as the $i'$th row in the 2D tensor ${\phivec}_{i'j'}$, and $\mathbf{K}\in\mathbb{R}^{M_\text{d}\times M_\text{d}}$ as the finite-difference operator kernel, we can express these operations as $\phivec_{i':}'=\phivec_{i',:} \mathbf{K}$. For illustrative purpose, consider the central-difference discretization of $d_\xi\phi$ on an equidistant mesh with $M_\text{d}$ grid points, grid size $\Delta$, and periodic boundaries, the discrete form can be written as:
\begin{equation}
 \phivec_{i',:} = (\phi_1, \phi_2, \ldots, \phi_{M_\text{d}-1}, \phi_{M_\text{d}})
\end{equation}
and
\begin{equation}
 \mathbf{K}= \f{1}{2\Delta}\begin{pmatrix}
 0 & -1 & 0 & 0 & \cdots  & 0 & 0\\
 1 &  0 & -1 & 0 &\cdots  & 0 & 0\\
 \vdots &\vdots&\vdots&\vdots&\ddots&\vdots&\vdots\\
  0 & 0 &  0 & 0 & \cdots & 0 & -1\\
  0 & 0 &  0 & 0 & \cdots & 1 & 0\\
  \end{pmatrix}\;.
\end{equation}

\begin{figure}[!htb!]
    \centering
    \includegraphics[width=0.6\textwidth]{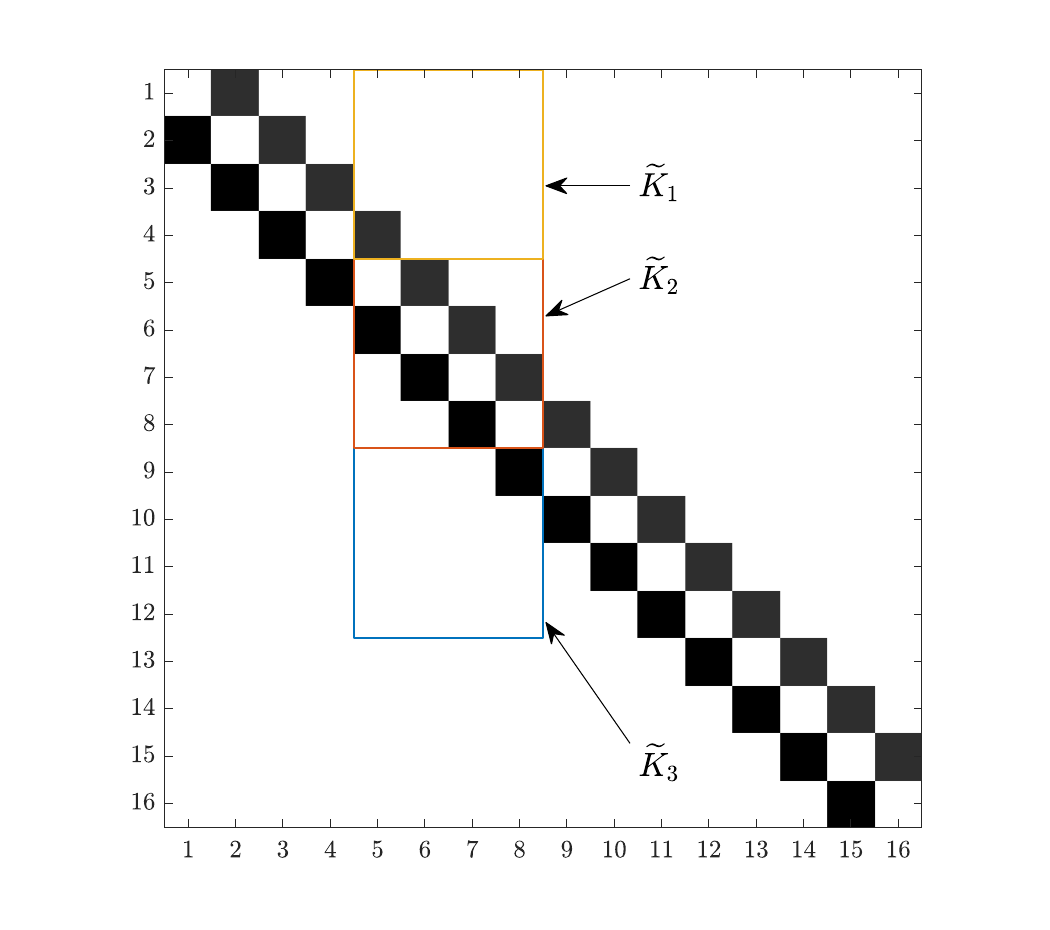}
    \caption{Illustration of finite-difference kernel $\mathbf{K}$ applied to a vector of length $M_\text{w}=16$. The sub-block size of the convolution kernel $\wt{\mathbf{K}}_l$ is $M_\text{c}=4$.}
    \label{fig:conv_kernel}
\end{figure}

\begin{figure}[!htb!]
    \centering
    \includegraphics[width=0.8\textwidth]{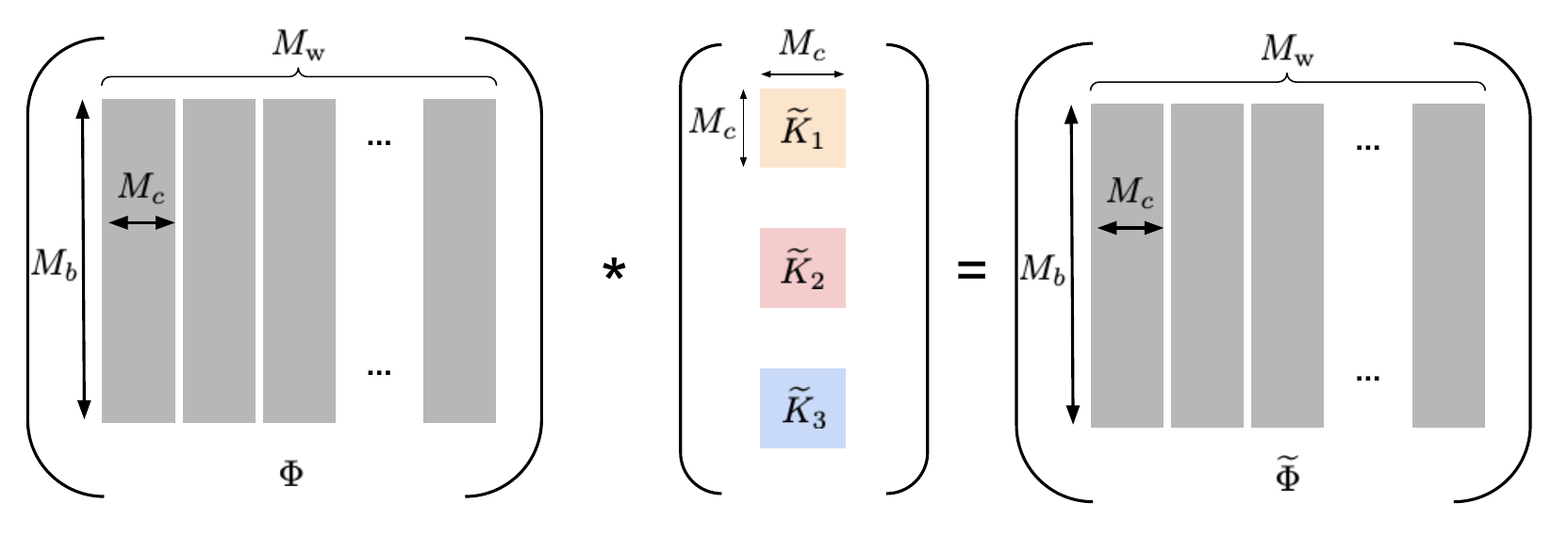}
    \caption{Graphic representation of the \lstinline{tf.nn.conv1d} function that is used to perform a discretized operation.}
    \label{fig:conv_explanation}
\end{figure}

By taking advantage of the sparsity of $\mathbf{K}$, we replace the vector-matrix multiplication with a convolution using \lstinline{tf.nn.conv1d}. The convolution kernel is constructed from three consecutive sub-blocks in $\mathbf{K}$, which is denoted as $\wt{\mathbf{K}}=[\wt{\mathbf{K}}_1,\wt{\mathbf{K}}_2,\wt{\mathbf{K}}_3]$ where the size of $\wt{\mathbf{K}}_l$ is customizable through the specification of $M_\text{c}$. The channel width $M_\text{c}$ is a parameter satisfying these conditions:
\begin{itemize}
    \item $3\times M_\text{c} $ is greater or equal to the stencil width of the discrete operator;
    \item The physical dimension of the input \lstinline{tf.Tensor} should be greater than the channel size;
    \item The physical dimension of the input \lstinline{tf.Tensor} should be fully divisible by the channel size.
\end{itemize}
For different physical dimensions, we can adjust the size of the sub-blocks to achieve optimal performance. To leverage the TPU-specific hardware architecture to accelerate matrix operations, $M_\text{c}$ should be a multiple of 8. Other channel sizes lead to undesired zero-padding by the XLA compiler, hence results in computational inefficiencies. For illustration, \Cref{fig:conv_kernel} shows an example of sub-blocks with $M_\text{c}=4$ and $M_\text{w}=4$, which provides $M_\text{d}=16$. To accommodate this data structure, we reshape $\phivec_{i',:}$ into a matrix $\Phi_{i^\prime}\in\mathbb{R}^{M_\text{c}\times M_\text{w}}$ in row-major order. With this transformation, the original plane $\phivec_{i'j'}$ becomes $\Phi\in \mathbb{R}^{M_\text{b}\times M_\text{c}\times M_\text{w}}$. The three dimensions in $\Phi$ correspond to the batch size, channels, and input width, which conforms with the arguments of \lstinline{tf.nn.conv1d}. The output of this function provides the result of the finite-difference operator as $\wt{\Phi}=\Phi * \wt{K}$, which is considered as block convolutions and illustrated graphically in~\cref{fig:conv_explanation}. This convolution operation is applied to all $\phi_{ij}$ repeated over $k\in[1,\wt{N}_z]$. 

\begin{figure}[!htb!]
 \centering
 \includegraphics[width=0.5\columnwidth]{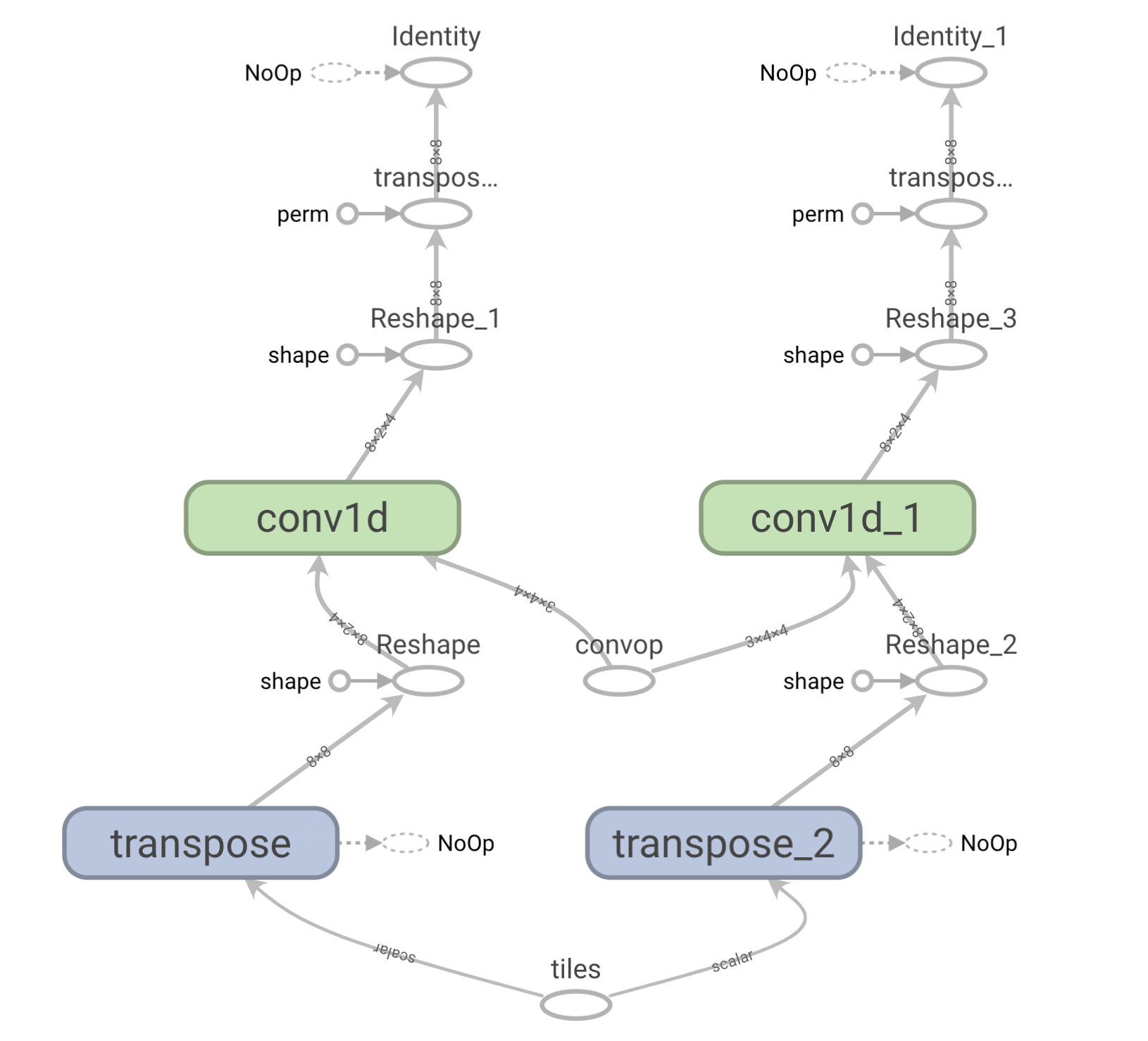}
 \caption{\label{fig:kernel_op_graph_x}Tensorflow graphs corresponding to kernel operation for finite-difference operation along $x$-direction in~\cref{lst:kernel_x}.}
\end{figure}

\Cref{lst:kernel_x} shows the implementation of the convolution operation in the $x$-direction, and~\cref{fig:kernel_op_graph_x} shows an example of the TensorFlow graph that corresponds to this implementation. Convolution in the $y$-direction is implemented in a similar form. In this implementation, \lstinline{tiles} is the 3D data structure to which the finite-difference operation is applied, with each \lstinline{tile} being an $x$-$y$ plane. The same kernel operation is applied to all $x$-$y$ planes along the $z$-direction. In this particular example, $\wh{N}_z = 2$ and the operations are represented by two identical branches. In each branch, the 2D \lstinline{tf.Tensor} is first reshaped by the \lstinline{transpose} and \lstinline{reshape} operations into a 3D \lstinline{tf.Tensor} that conforms to the function \lstinline{tf.nn.conv1d}, as described above. The transformed tensors from different branches are subsequently convolved with the same kernel \lstinline{convop} that defines the finite-difference operation. The result of the \lstinline{conv1d} operation is finally reshaped back to a 2D \lstinline{tf.Tensor}. Combining results from all branches completes the function.

One point to note in~\cref{lst:kernel_x} is that the output tensor adopts the same shape as the input only if the padding mode in the \lstinline{tf.nn.conv1d} function is set to \lstinline{'SAME'}. In this mode, TensorFlow assumes that 2 column blocks with zeros of size $M_\text{b}\times M_\text{c}$ will be appended to the two ends of the input tensor, which allows the convolution to be conducted normally. If the padding mode is set to \lstinline{'VALID'}, no augmentation is done for the input tensor, and the convolution operations are only performed for points that are valid. As a result, the number of output blocks is $M_\text{w} - 2$. To preserve the size of the computational mesh, we use \lstinline{'SAME'} as the padding mode in \lstinline{tf.nn.conv1d}. Note that boundary condition enforcement is required after applying kernel operations.

\begin{lstlisting}[float=tp,floatplacement=tbp,language=Python,basicstyle=\scriptsize,linerange={1-15},caption={Kernel function that is used to performs finite difference operations along the x-direction.},captionpos=b,label={lst:kernel_x}]
def apply_convolutional_op_x(tiles: Iterable[tf.Tensor],
                             convop: tf.Operation) -> List[tf.Tensor]:
  """Applies kernel operations in the x direction."""
  kernel_size = convop.shape.as_list()
  result = []
  for tile in tiles:
    x_size, y_size = tile.shape.as_list()
    reshaped_transposed_input = tf.reshape(
        tf.transpose(tile, [1, 0]), [y_size, -1, kernel_size[-1]])
    convolved_output = tf.nn.conv1d(
        reshaped_transposed_input, convop, 1, 'SAME')
    reshaped_output = tf.transpose(
        tf.reshape(convolved_output, [y_size, -1]), [1, 0])
    result.append(reshaped_output)
  return result
\end{lstlisting}

Finite-difference operations along the $z$-direction are performed by slicing through the 2D planes, as shown in~\cref{lst:kernel_z}. In this implementation, $x$-$y$ planes are combined linearly by a series of \lstinline{saxpy} operations. For $k\in[\lfloor S/2\rfloor, \wh{N}_z - \lfloor S/2\rfloor]$, the kernel operations is performed as:
\begin{equation}
    \widetilde{\Phi}_{k}=\sum_{s=-\lfloor S/2\rfloor}^{\lfloor S/2\rfloor}\Phi_{k+s} K_s,
    \label{eq:saxpy}
\end{equation}
where $S$ is the width of the stencil of the scheme. For $k<\lfloor S/2\rfloor$ and $k>N_z-\lfloor S/2\rfloor$, no kernel operation is applied, and $\widetilde{\Phi}_{k}=\Phi_{k}$. These cells are considered as ghost cells. The values in these cells are updated with boundary conditions.

\begin{lstlisting}[float=tp,floatplacement=tbp,language=Python,basicstyle=\scriptsize,linerange={17-32},caption={Kernel function that is used to performs finite difference operations along the z-direction.},captionpos=b,label={lst:kernel_z}]
def apply_op_z(tile_list: Sequence[tf.Tensor],
               z_op_list: Sequence[tf.Operation],
               shift: Sequence[int] = None) -> List[tf.Tensor]:
  """Applies kernel operations in the z direction."""
  start_shift = min(shift)
  end_shift = max(shift)
  out_list = []
  range_start = max(0, -start_shift)
  range_end = min(len(tile_list) - end_shift, len(tile_list))
  for i in range(range_start, range_end):
    out = tile_list[i + shift[0]] * z_op_list[0]
    for j in range(1, len(shift)):
      out += tile_list[i + shift[j]] * z_op_list[j]
    out_list.append(out)
  return ([tile_list[i] for i in range(range_start)] + out_list +
          [tile_list[i] for i in range(range_end, len(tile_list))])
\end{lstlisting}

To illustrate how the kernel operations are employed, we discuss the specific implementation of the QUICK scheme (see~\cref{SSEC_NUM_DISCRETIZATION}). \Cref{lst:quick} illustrates the computation of the  convection term. In this implementation, the weights of the finite-difference operation are stored in a kernel-operation dictionary. The keys in the dictionary are the names of operators, and the corresponding values are the weights of these operations. In~\cref{lst:quick}, for example, the kernel operation that computes face fluxes with a positive face velocity is named as \lstinline{'kf2-'}, and the weights of this operator are $(-0.125$, $0.75$, $0.375)$. Operations with specific names are performed with kernel functions defined in~\cref{lst:kernel_x,lst:kernel_z}. The convective flux is then obtained by performing a series of \lstinline{saxpy} operations for $x$-$y$ planes at the same location in the intermediate 3D tensors, as shown in the \lstinline{return} statement from line 28 to 32 in~\cref{lst:quick}, which is a discrete representation of~\cref{eq:quick_flux}. Note that the \lstinline{saxpy} operations are performed for all $x$-$y$ planes by looping over the $z$-direction in a list. Following this, the convection term is computed by applying a first-order gradient to the flux between neighboring nodes with another kernel operation, as shown by the \lstinline{return} statement in line 53 of~\cref{lst:quick}. 

\begin{lstlisting}[float=tp,floatplacement=tbp,language=Python,basicstyle=\scriptsize,linerange={34-86},caption={Implementation of the QUICK scheme that is used for computing the convection terms.},captionpos=b,label={lst:quick}]
def face_flux_quick(state: Sequence[tf.Tensor], rhou: Sequence[tf.Tensor],
                    pressure: Sequence[tf.Tensor], dx: float, dt: float,
                    dim: int) -> List[tf.Tensor]:
  """Computes the face flux of `state` normal to `dim` with QUICK scheme."""
  kernel_op = get_kernel_fn.ApplyKernelConvOp(
      4, {'kf2-': ([-0.125, 0.75, 0.375], 2)})

  if dim == 0:
    quick_pos_type = ['kf2-x']
    quick_neg_type = ['kf2x+']
    kernel_fn = kernel_op.apply_kernel_op_x
  elif dim == 1:
    quick_pos_type = ['kf2-y']
    quick_neg_type = ['kf2y+']
    kernel_fn = kernel_op.apply_kernel_op_y
  elif dim == 2:
    quick_pos_type = ['kf2-z', 'kf2-zsh']
    quick_neg_type = ['kf2z+', 'kf2z+sh']
    kernel_fn = kernel_op.apply_kernel_op_z
  else:
    raise ValueError('`dim` has to be 0, 1, or 2. {} is provided.'.format(dim))

  state_pos = kernel_fn(state, *quick_pos_type)
  state_neg = kernel_fn(state, *quick_neg_type)
  rhou_face = face_interpolation(kernel_op, rhou, pressure, dx, dt, dim)

  # Notations mapped to Eq. (3.17): rhou->m_j, s_pos->\phi^+, s_neg->\phi^-
  return [
      0.5 * (rhou_i + tf.abs(rhou_i)) * s_pos_i + 0.5 *
      (rhou_i - tf.abs(rhou_i)) * s_neg_i
      for rhou_i, s_pos_i, s_neg_i in zip(rhou_face, state_pos, state_neg)
  ]
  
def convection_quick(kernel_op: get_kernel_fn.ApplyKernelOp,
                     state: Sequence[tf.Tensor], rhou: Sequence[tf.Tensor],
                     pressure: Sequence[tf.Tensor], dx: float, dt: float,
                     dim: int) -> List[tf.Tensor]:
  """Compute the convection term for convservative variables using QUICK scheme."""
  if dim == 0:
    diff_op_type = ['kdx+']
    kernel_fn = kernel_op.apply_kernel_op_x
  elif dim == 1:
    diff_op_type = ['kdy+']
    kernel_fn = kernel_op.apply_kernel_op_y
  elif dim == 2:
    diff_op_type = ['kdz+', 'kdz+sh']
    kernel_fn = kernel_op.apply_kernel_op_z
  else:
    raise ValueError('`dim` has to be 0, 1, or 2. {} is provided.'.format(dim))

  flux = face_flux_quick(state, rhou, pressure, dx, dt, dim)

  return [d_flux / dx for d_flux in kernel_fn(flux, *diff_op_type)]
\end{lstlisting}

\section{\label{SEC_SINGLE_PRECISION_ARITH}Single-precision arithmetic}
While scientific computing applications are commonly performed using double-precision arithmetic, less precision is typically required by deep-learning applications. In the interest of speed, power usage, and silicon area of the MXU, the TPU instruction sets utilize bfloat16 and float32 arithmetic for its MXU. Specifically, the MXU array performs multiplications in bfloat16, and accumulation with accuracy equivalent to float32. Additionally, based on this hardware, we can emulate float32 in software. The software-based emulation makes 32-bit floating point arithmetic operations available with a performance trade-off. The results presented in this work are all performed with 32-bit floating point arithmetic.

Numerical errors arising from the reduced floating-point accuracy in comparison to typical 64-bit arithmetic can contaminate the numerical solution. Because of its relevance for high-fidelity fluid-flow simulations on TPU architectures, we examine effects of floating-point accuracy and provide guidelines for mitigating these errors. For this analysis, we consider the following scalar advection-diffusion problem in one dimension:
\begin{equation}
 \label{EQ_ADE_MODEL} 
    \p_t\phi + u\p_x \phi = \alpha \p_{xx}\phi\;,
\end{equation}
for $x\in[0,1]$ with periodic boundary conditions, and constant velocity $u$ and diffusivity $\alpha$. An analytic solution to~\cref{EQ_ADE_MODEL} is given as:
\begin{equation}
 \label{EQ_ADE_EXACT}
 \varphi(x,t) = \sin(2\pi [x-ut])\exp\left\{- \frac{t}{t_c}\right\}\;,
\end{equation}
with the characteristic decay rate $t_c=(4\pi^2\alpha)^{-1}$.
We discretize the temporal derivative with a first-order forward Euler scheme and all spatial operators are discretized using a second-order finite-difference scheme, resulting in the following discrete form:
\begin{equation}
 \label{EQ_ADE_FD}
    \f{\phi^{n+1}_i-\phi^{n}_i}{\Delta t} = 
    - u\f{\phi^{n}_{i+1}-\phi^{n}_{i-1}}{2\Delta x} 
    + \alpha \f{\phi^{n}_{i+1}-2\phi^{n}_{i}+\phi^{n}_{i-1}}{\Delta x^2}.\;
\end{equation}
To analyze the error that arises from the floating-point representation, we expand all quantities in~\cref{EQ_ADE_FD} as follows:
\begin{equation}
 \label{EQ_ADE_FD_FLOATING}
 \begin{split}
    \f{(\phi^{n+1}_i+\epsilon_1)-(\phi^{n}_i+\epsilon_2)}{\Delta t+\tau} = &
    - u\f{(\phi^{n}_{i+1}+\epsilon_3)-(\phi^{n}_{i-1}+\epsilon_4)}{2(\Delta x+\delta_1)} \\
    &+ \alpha \f{(\phi^{n}_{i+1}+\epsilon_5)-2(\phi^{n}_{i}+\epsilon_6)+(\phi^{n}_{i-1}+\epsilon_7)}{(\Delta x+\delta_2)(\Delta x+\delta_3)},\;
  \end{split}
\end{equation}
where the round-off errors $\epsilon_i, \tau, \delta_j\sim{\cal{P}}(0,\sigma)$ are sampled from a distribution $\cal{P}$ with zero-mean and a standard deviation $\sigma$ that depends on the compute architecture and precision. By retaining first-order terms in the round-off error,~\cref{EQ_ADE_FD_FLOATING} can be written as:
\begin{equation}
 \label{EQ_ADE_FD_FLOATING_EXP}
 \begin{split}
    \f{\phi^{n+1}_i-\phi^{n}_i}{\Delta t} = 
    - u\f{\phi^{n}_{i+1}-\phi^{n}_{i-1}}{2\Delta x} 
    + \alpha \f{\phi^{n}_{i+1}-2\phi^{n}_{i}+\phi^{n}_{i-1}}{\Delta x^2}+E^n_{\text{RO},i},\;
  \end{split}
\end{equation}
with the pointwise round-off error, $E^n_{\text{RO},i}$, taking the following form:
\begin{equation}
 \label{EQ_ADE_FD_FLOATING_EXP_EROE}
  \begin{split}
 E^n_{\text{RO},i} = &-\f{1}{\Delta t}\left[\epsilon_1-\epsilon_2 - (\phi^{n+1}_i-\phi^{n}_i)\f{\tau}{\Delta t}\right]\\
                      &-\f{u}{2\Delta x}\left[\epsilon_3 - \epsilon_4 - (\phi^{n}_{i+1} - \phi^n_{i-1})\f{\delta_1}{\Delta x}\right]\\
                      &+\f{\alpha}{\Delta x^2}\left[\epsilon_5 - 2\epsilon_6+\epsilon_7 - 
                                                    (\phi^{n}_{i+1} - 2\phi^n_{i} + \phi^{n}_{i-1})\left(\f{\delta_2}{\Delta x}+\f{\delta_3}{\Delta x}\right)\right].
 \end{split}
\end{equation}
The local error can then be computed by subtracting~\cref{EQ_ADE_EXACT} from~\cref{EQ_ADE_FD_FLOATING_EXP}, giving:
\begin{equation}
\begin{split}
\label{EQ_ERROR_ANALYTIC}
    e_i = \phi^{n+1}_i-\varphi^{n+1}_i = -\f{u}{6}{\varphi'''(x_i)}\Delta t\Delta x^2 + \f{\alpha}{12}\varphi''''(x_i)\Delta t\Delta x^2 -\f{1}{2}\ddot{\varphi}(x_i)\Delta t^2+ \Delta t E^n_{\text{RO},i},
\end{split}
\end{equation}
where the first three terms on the right-hand side represent the truncation errors for the approximation of the spatial and temporal derivatives. 

\begin{figure}[!htb!]	
 \centering
 \begin{subfigure}[t]{0.49\columnwidth}
  \centering
  \includegraphics[width=\columnwidth]{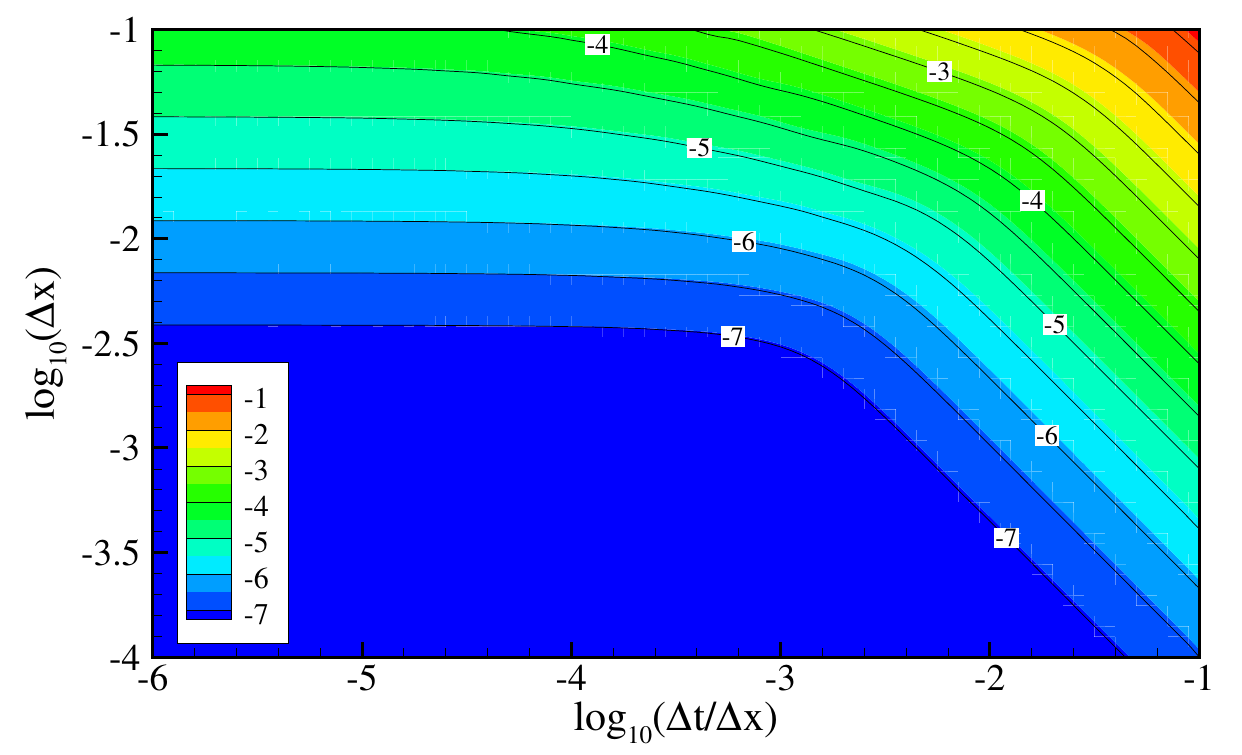}
  \caption{\label{FIG_ERROR_ANALYSIS_DP}Double precision arithmetic.}		
 \end{subfigure}
 \hfill
 \begin{subfigure}[t]{0.49\columnwidth}
  \centering
  \includegraphics[width = \columnwidth]{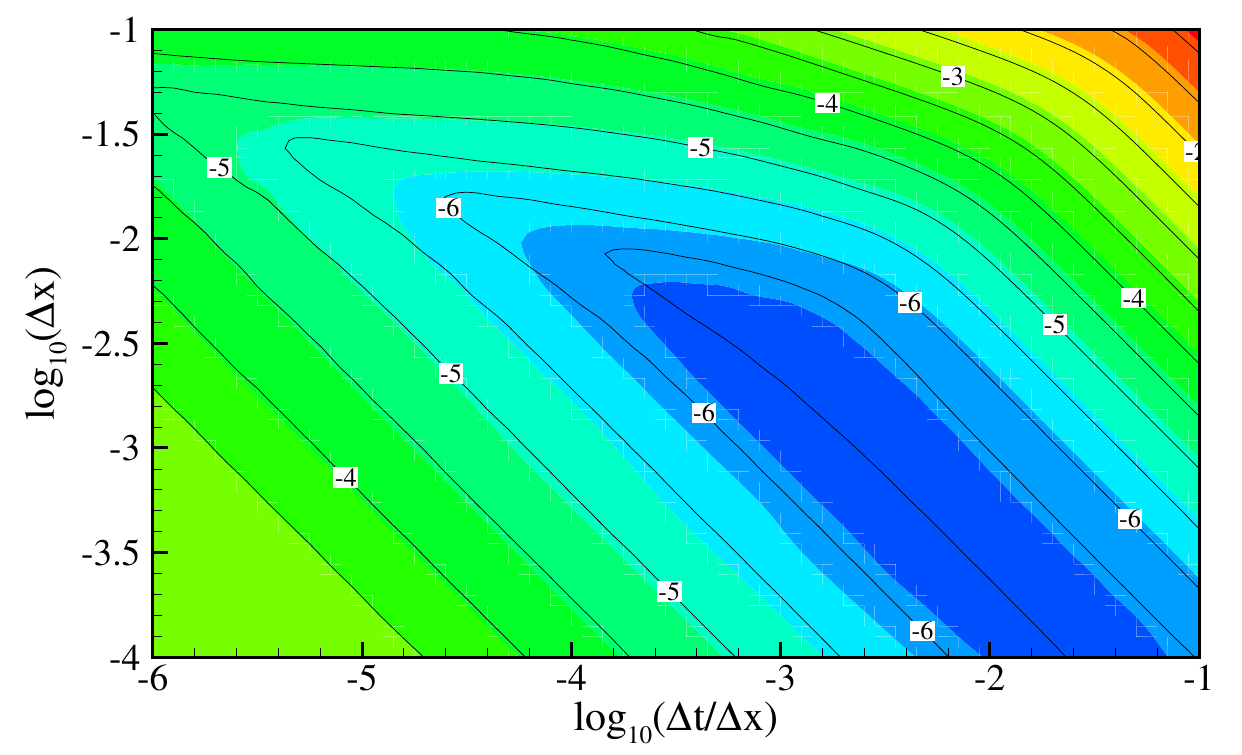}
  \caption{\label{FIG_ERROR_ANALYSIS_SP}Single precision arithmetic.}		
 \end{subfigure}
 \caption{\label{FIG_ERROR_ANALYSIS_DP_SP}$L_1$ error norm (on logarithmic scale) combining truncation and round-off error using (a) double-precision and (b) single-precision floating point accuracy. The colored isosurface shows the numerical error, $\|\phi-\varphi\|,$ and isolines show the analytic error evaluated from the right-hand side of~\cref{EQ_ERROR_ANALYTIC}.}
 \end{figure}
A comparison of errors from the numerical simulation and the analytic solution in~\cref{EQ_ERROR_ANALYTIC}, evaluated as $L_1$-norm, is shown in~\cref{FIG_ERROR_ANALYSIS_DP_SP}. The axis are chosen to demonstrate the direct dependence on the convective CFL number. The simulations are performed using $u=5$ and $\alpha=5\times10^{-5}$. The solution is advanced until $10^{-7}t_c$. The results are shown as a function of spatial and temporal resolution. Isocontours illustrate the numerical error and isolines correspond to the analytic error, given by~\cref{EQ_ERROR_ANALYTIC}. Simulations performed using double-precision arithmetic (\cref{FIG_ERROR_ANALYSIS_DP}) show the expected behavior of the truncation error with reduction in mesh resolution and time-step size. In contrast, the error introduced by the round-off error competes with the truncation error with increasing spatial and temporal resolution for the single-precision arithmetic (\cref{FIG_ERROR_ANALYSIS_SP}). 

Given information about the accuracy of the floating point arithmetic employed, we can use~\cref{EQ_ERROR_ANALYTIC} to estimate the optimal resolution for numerical simulations. By retaining only leading-order terms in~\cref{EQ_ERROR_ANALYTIC} and assuming that all round-off errors are bounded by $\sigma$, the following error estimate is obtained:
\begin{equation}
\label{EQ_ERROR_ANALYTIC_ESTIMATE}
\|e\| = {\cal O}\left(\f{\Delta t}{\Delta x}\Delta x^3\right)+{\cal O}\left(\left(\f{\Delta t}{\Delta x}\right)^2\Delta x^2\right)  + {\cal O}\left(\f{\sigma}{\Delta x}\f{\Delta x}{\Delta t}\right)\;.
\end{equation}
This expression provides a relation for estimating the resolution requirement to minimize the error. Taking the derivative of~\cref{EQ_ERROR_ANALYTIC_ESTIMATE} with respect to $\Delta t$, it follows that 
\begin{equation}
 \Delta t \propto \begin{cases}\displaystyle
  \frac{{\sigma}^{1/2}}{\Delta x}\quad \text{for $\Delta t\ll\Delta x^2$}\\
  \sigma^{1/3}\quad \text{otherwise}\\
    \end{cases}\;,
\end{equation}
where the first condition is most relevant for advection-dominated problems. With this, $\sigma$ can be estimated from the machine precision with $\sigma=2^{-23}$ for single precision and $\sigma=2^{-52}$ for double precision. 

This analysis illustrates the importance of considering optimal resolution requirements so that the solution is minimally impacted by round-off errors when using single-precision arithmetic.

Techniques such as the compensated summation~\cite{KAHN_1965,Higham_SIAM1993} could be employed to mitigate round-off errors. This was not explored further in this work. We also note that 64-bit floating point operations are supported at the software level on 
 TPUs~\cite{HENRY_TANG_HEINECKE_ARXIV2019}. However, the algorithmic complexity impacts run-time performance and was not employed in this work. In the following, we will examine the convergence and numerical accuracy of single-precision floating point arithmetic for fluid-flow simulation on TPU-instruction sets.
\section{\label{SEC_RESULTS}Results}
This section is concerned with verifying the implementation of the TPU-algorithm and assessing the performance on TPU architectures. To this end, we consider four different flow configurations that include Taylor-Green vortex (TGV) flows in two and three dimensions, homogeneous isotropic turbulence (HIT), and a turbulent planar jet. This is complemented by examining the parallel code performance by considering weak and strong scalability in~\cref{SEC_PERFORMANCE}.
\subsection{\label{SSEC_2DTGV}2D Taylor-Green vortex flow}
The first test case under consideration is the 2D TGV flow, which has been established as a CFD benchmark. Analysis of the TGV-flow has shown that the solution is susceptible to perturbations arising from truncation errors and round-off errors~\cite{SENGUPTA_SHARMA_SENGUPTA_POF2018}, making this case particularly interesting for examining convergence and solution accuracy of our TPU implementation. Despite of instabilities induced by perturbations, an analytical solution is available for the base flow of this spatio-temporally evolving flow.

The analytic solution of the 2D TGV flow can be written as~\cite{MEI_LUO_LALLEMAND_ETAL_CF2006}:
\begin{subequations}
\label{EQ_2D_TGV}
\begin{align}
    u(\boldsymbol{x},t) &= U_0\sin \left(\frac{ x}{L}\right)\cos \left(\frac{ y}{L}\right)\exp\left\{-\frac{t}{t_c}\right\}, \\
    v(\boldsymbol{x},t) &= -U_0\cos \left(\frac{ x}{L}\right)\sin \left(\frac{ y}{L}\right)\exp\left\{-\frac{t}{t_c}\right\}, \\
    p(\boldsymbol{x},t) &= \frac{\rho_0 U_0^2}{4} \left[\cos \left(2\frac{ x}{L}\right)+\cos \left(2\frac{ y}{L}\right)\right]\exp\left\{-2\frac{t}{t_c}\right\}\,.
\end{align}
\end{subequations}
The computational domain consists of a square with $\boldsymbol{x}\in[0,2\pi L]^2$, with $L=1\,\text{m}$ being the characteristic length. The characteristic time scale is defined with respect to the viscous dissipation, giving 
$t_c=L^2/(2\nu)$. Periodic boundary conditions are applied along both directions. In this simulation, fluid properties are assumed to be constant. The kinematic viscosity is set to $\nu = 6.25\times10^{-4}\ \text{m}^2/\text{s}$, resulting in a specific Reynolds number of  $Re=1/\nu=1600$~\cite{Van_Rees2011-jz}. All other quantities are unity, i.e., $U_0=1\ \text{m/s}$ and $\rho_0=1\ \text{kg/m}^3$. 

The computational domain is discretized using an equidistant mesh with $N=L/\Delta$ grid points in each direction, where $\Delta$ is the isotropic grid spacing. To examine the  effect of the spatial and temporal resolution on the convergence, simulations are either performed at a constant time-step size of $\Delta t=2\times 10^{-3}\,\text{s}$, or for a constant convective CFL number with $\text{CFL}=0.1$ and 0.3. All simulations are performed with mesh resolutions ranging from $N=32$ to $N=1024$, and the solutions are advanced until $t=0.05t_c$. The results are compared with the analytical solution at the end of the simulation.

\begin{figure}[!htb!]
  \centering
  \begin{subfigure}[t]{0.49\columnwidth}
    \centering
    \includegraphics[width=\textwidth]{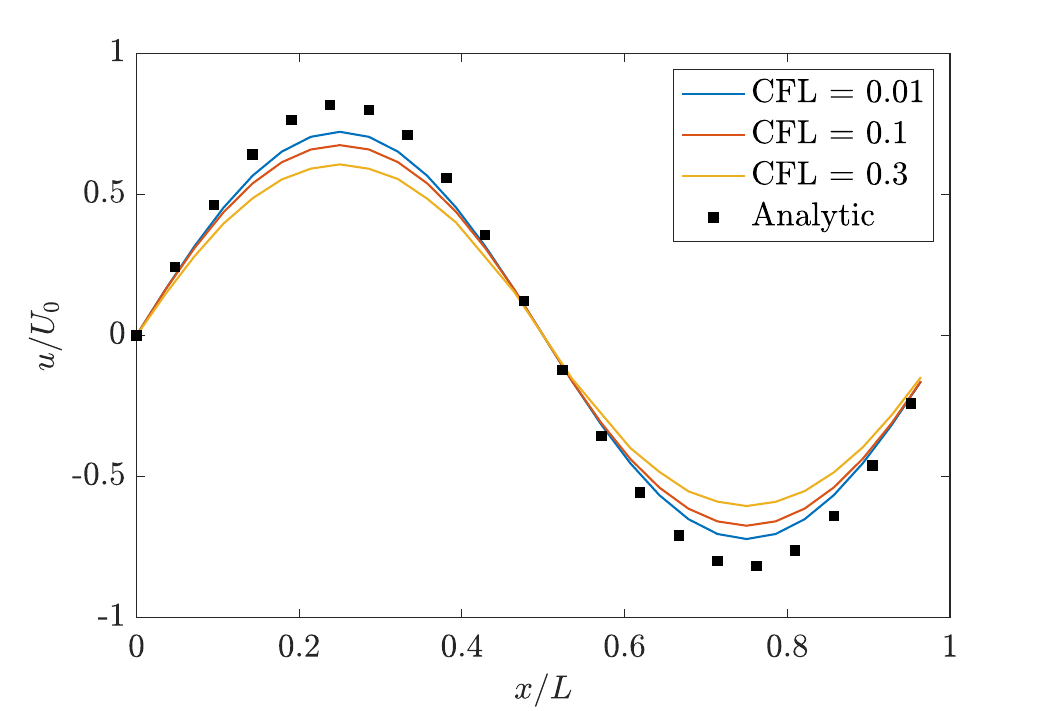}
    \caption{\label{fig:tgv2d_yL05_N32}$N=32$.}
  \end{subfigure}
  \hfill
  \begin{subfigure}[t]{0.49\columnwidth}
    \centering
    \includegraphics[width=\textwidth]{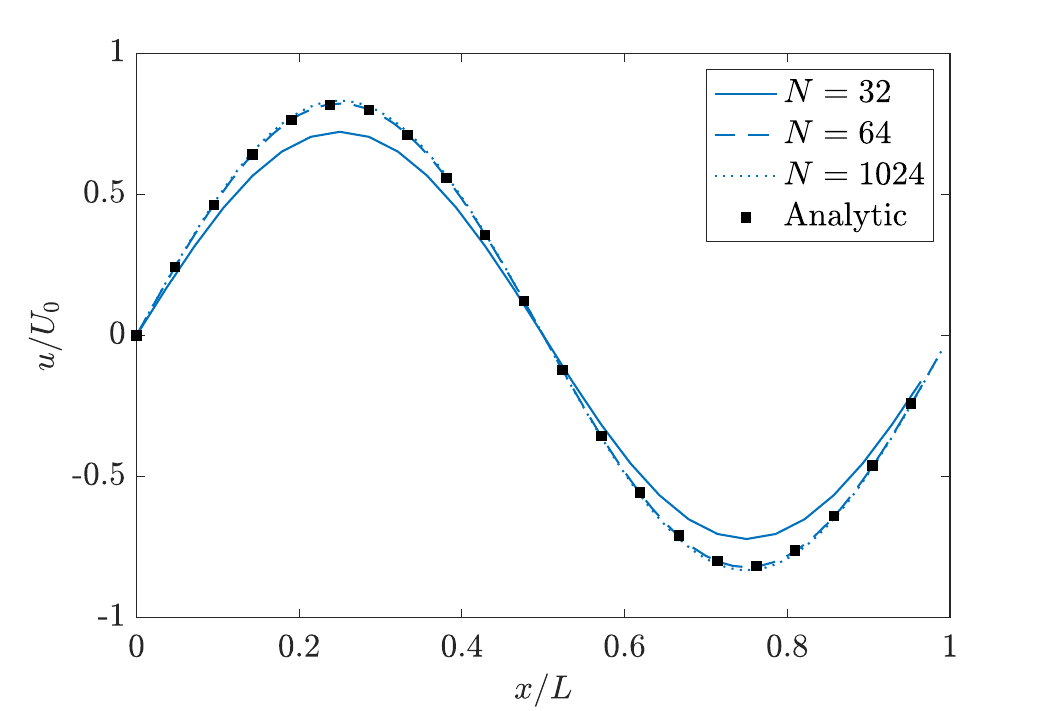}
    \caption{\label{fig:tgv2d_yL05_N32}$\Delta t=0.002$ s.}
  \end{subfigure}
  \caption{ \label{FIG_2DTGV_CENTERLINE}Comparison of instantaneous velocity $u$ for different mesh resolutions with analytic solution at $y/L=0.08$ and $t=0.05t_c$; (a) Variation in CFL number for fixed mesh resolution of $N=32$; (b) Variation in mesh resolution for a fixed time step size $\Delta t =0.002$ s.}
\end{figure}
\Cref{FIG_2DTGV_CENTERLINE} shows comparisons of instantaneous velocity profiles for three different mesh resolutions with the analytic results of~\cref{EQ_2D_TGV}. The velocity profiles are extracted at $y/L=0.08$. Results shown by the solid curves are computed with a mesh resolution of $N=32$, and it can be seen that the deviation from the analytic solution increases with increasing CFL number (note that for this case, $\Delta t=0.002\,\text{s}$ corresponds to a CFL number of $0.01$). However, with increasing spatial resolution, the instantaneous solution converges to the analytic solution, as shown by the results for $N=64$ and $N=1024$, which were obtained for $\Delta t=0.002\,\text{s},$ corresponding to a CFL number of $0.02$ and $0.326,$ respectively.

\begin{figure}[!htb!]
  \centering
  \includegraphics[width=0.55\textwidth]{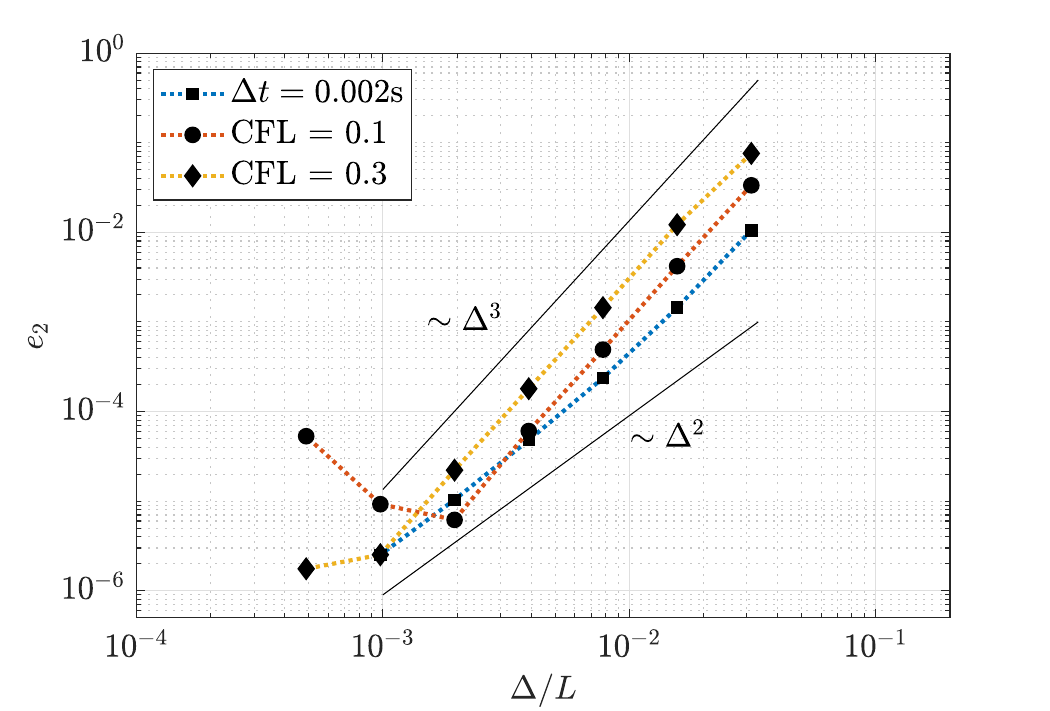}
  \caption{\label{FIG_2DTGV_L2_NORM}Convergence plot showing $L_2$-norm of velocity 
                                    magnitude as a function of normalized grid spacing $\Delta /L$.}
\end{figure}

To quantify the convergence with respect to the mesh resolution, we compute the $L_2$-error norm of the velocity as:
\begin{equation}
 \label{EQ_L2_ERROR}
 e_2^2(t) = \iint \left[\boldsymbol{u}(\boldsymbol{x},t)-\boldsymbol{u}_{\text{analyt}}(\boldsymbol{x},t)\right]^2 d\boldsymbol{x}\;,
\end{equation}
where $\boldsymbol{u}_{\text{analyt}}$ corresponds to the analytic solution in~\cref{EQ_2D_TGV}. \Cref{FIG_2DTGV_L2_NORM} shows the rate of convergence for the velocity norm as a function of mesh resolution. It can be seen that, depending on the temporal resolution, a convergence rate between second order and third order is observed. Furthermore, it can be seen that with reduced CFL number, the contribution by the round-off errors increases, which is manifested by the increasing error for $N>512$ at $\text{CFL}=0.1$. These results are consistent with the analysis presented in~\cref{SEC_SINGLE_PRECISION_ARITH}, demonstrating that the criterion of~\cref{EQ_ERROR_ANALYTIC_ESTIMATE} provides a useful estimate for determining the resolution limits in order to mitigate round-off error.

\subsection{\label{SSEC_3DTGV}3D Taylor-Green vortex flow}
By extending the configuration discussed in the previous section, we consider the 3D Taylor-Green vortex flow as the second configuration. Unlike the 2D-counterpart, this flow does not have an analytic solution to describe the temporal dynamics. After an initial transition phase, the flow generates a cascade of vortical structures at increasingly finer scales that eventually decay by numerical dissipation. The initial condition is described by the following equations~\citep{Van_Rees2011-jz}:
\begin{subequations}
\begin{align}
    u_0(\boldsymbol{x})&=U_0\sin\left(2\pi \frac{x}{L}\right)\cos\left(2\pi\frac{y}{L}\right)\cos\left(2\pi\frac{z}{L}\right), \\
    v_0(\boldsymbol{x})&=-U_0\cos\left(2\pi\frac{ x}{L}\right)\sin\left(2\pi\frac{y}{L}\right)\cos\left(2\pi\frac{ z}{L}\right), \\
    w_0(\boldsymbol{x})&=0, \\
    p_0(\boldsymbol{x})&=\frac{\rho_0 U_0^2}{16}\left(\cos\left(\frac{4\pi x}{L}\right)+\cos\left(\frac{4\pi y}{L}\right)\right)\left(\cos\left(\frac{4\pi z}{L}\right)+2\right).
\end{align}
\end{subequations}
with $U_0=1\ \text{m/s}$ and $\rho_0=1\ \text{kg}/\text{m}^3.$ A cubic domain with side length $L=(2\pi)\,\text{m}$ is used, which corresponds to a characteristic length scale $L_c=L/2\pi=1$ m. Periodic boundary conditions are applied along all directions. The viscosity is set to $\nu=6.25\times 10^{-4}\,\text{m}^2/\text{s}$. Simulations are performed with mesh sizes of $N=\{128, 256, 512, 1024\}$ grid points in each direction. A constant time step size of $\Delta t = 2\times 10^{-3}\,\text{s}$ is used for all simulations, which corresponds to a convective CFL number of 0.3 for the case with the finest resolution. Defining the characteristic time scale as in~\cref{EQ_2D_TGV} with $t_c=L_c/U_0=1\,\text{s}$, the simulation is advanced until $t/t_c=20$.

\begin{figure}[!htb!]
  \centering
  \begin{subfigure}[t]{0.32\columnwidth}
    \centering
    {\includegraphics[clip,trim=15mm 0mm 0mm 8mm,width=\columnwidth]{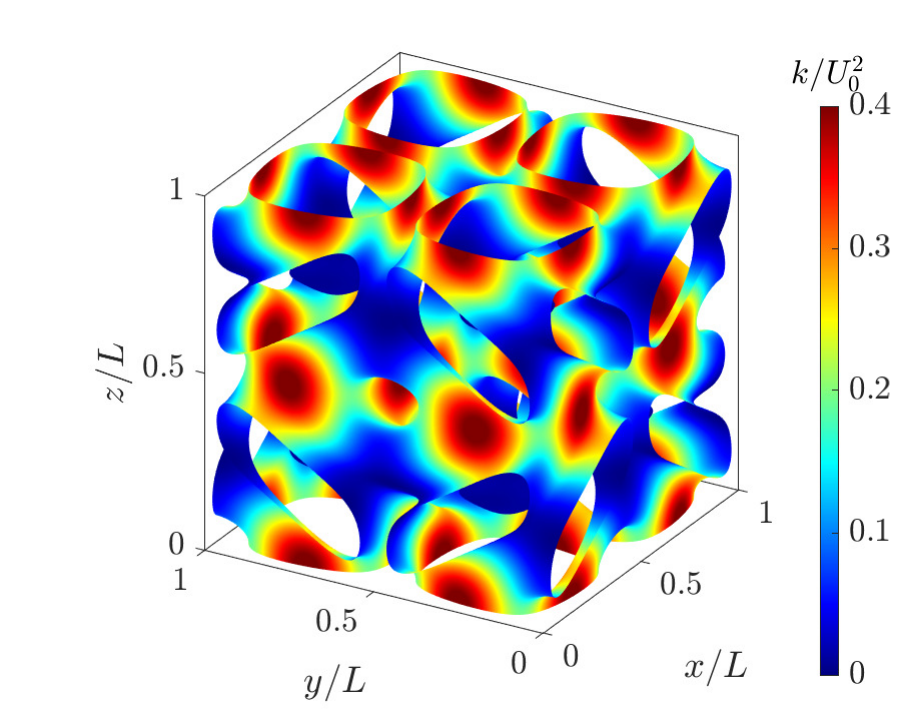}}
    \caption{\label{fig:tgv3d_vorticity_t1}$t/t_c=1$.}
  \end{subfigure}
  \hfill
  \begin{subfigure}[t]{0.32\columnwidth}
    \centering
    {\includegraphics[clip,trim=15mm 0mm 0mm 8mm,width=\columnwidth]{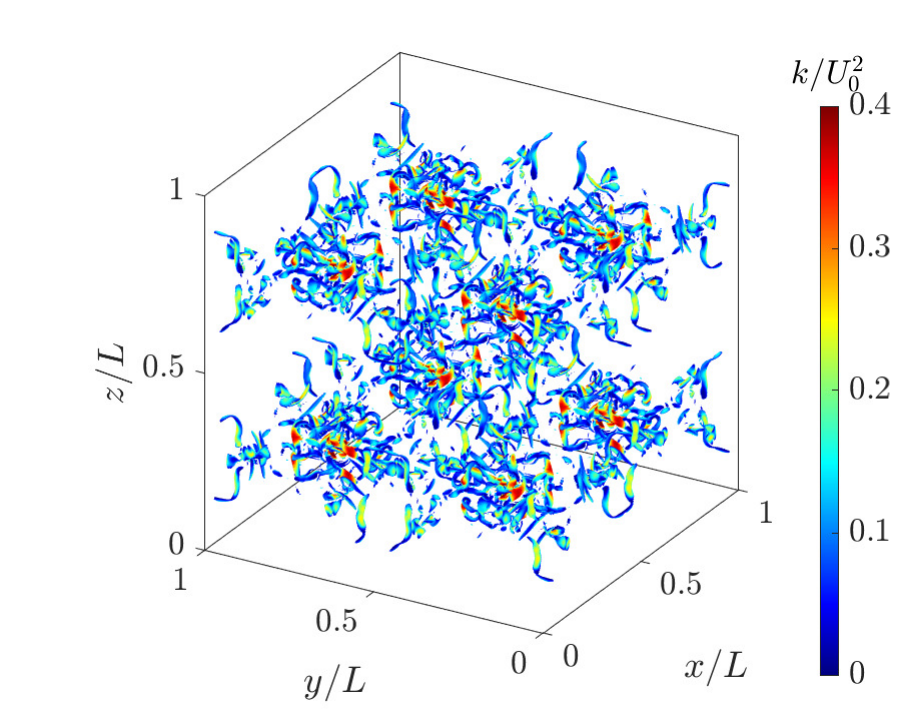}}
    \caption{\label{fig:tgv3d_vorticity_t10}$t/t_c=10$.}
  \end{subfigure}
  \hfill
  \begin{subfigure}[t]{0.32\columnwidth}
    \centering
    {\includegraphics[clip,trim=15mm 0mm 0mm 8mm,width=\columnwidth]{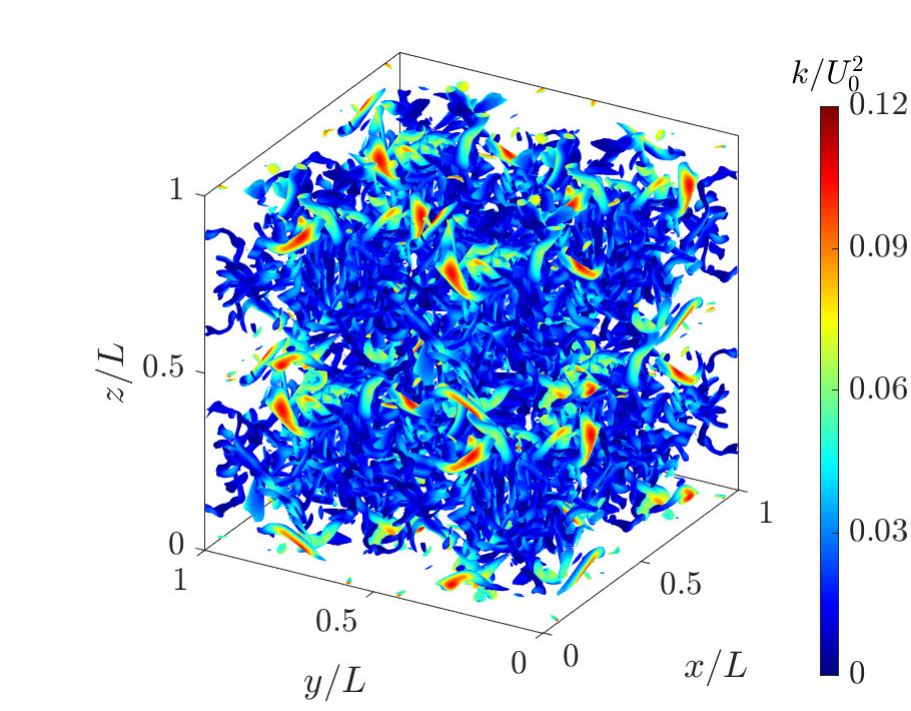}}
    \caption{\label{fig:tgv3d_vorticity_t20}$t/t_c=20$.}
  \end{subfigure}
  \caption{\label{fig:tgv3d_vorticity}Isosurface of vorticity magnitude $|\omegavec|/\max(|\omegavec|)=0.4$ colored by kinetic energy at (a) $t/t_c = 1$, (b) $t/t_c=10$, and (c) $t/t_c=20$.}
\end{figure}

Instantaneous 3D isosurfaces of vorticity with $|\omegavec|/\max(|\omegavec|)=0.4$ at three time instances are shown in~\cref{fig:tgv3d_vorticity}. The isosurfaces are colored by the normalized kinetic energy $k=|\uvec/U_0|^2/2$. These results illustrate the rapid decay of the initially coherent vortical structures into increasingly finer scales.

\begin{figure}[!htb!]
  \centering
  \begin{subfigure}[t]{0.49\columnwidth}
    \centering
    \includegraphics[width=\columnwidth]{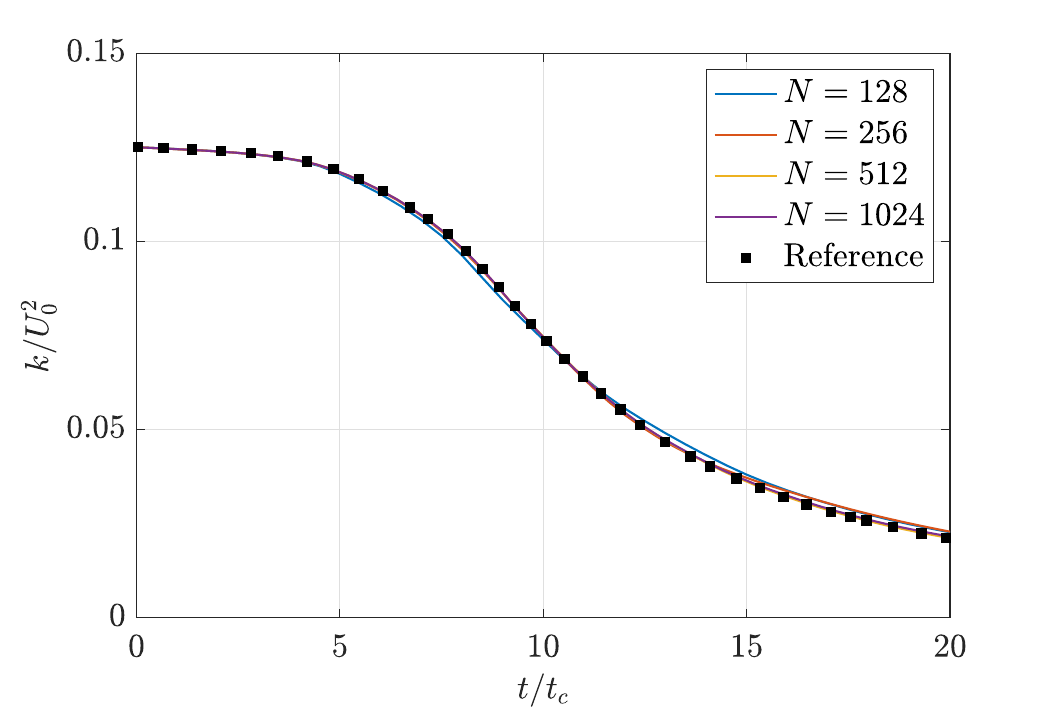}
    \caption{\label{fig:tgv3d_tke}Normalized kinetic energy.}
  \end{subfigure}
  \hfill
  \begin{subfigure}[t]{0.49\columnwidth}
    \centering
    \includegraphics[width=\columnwidth]{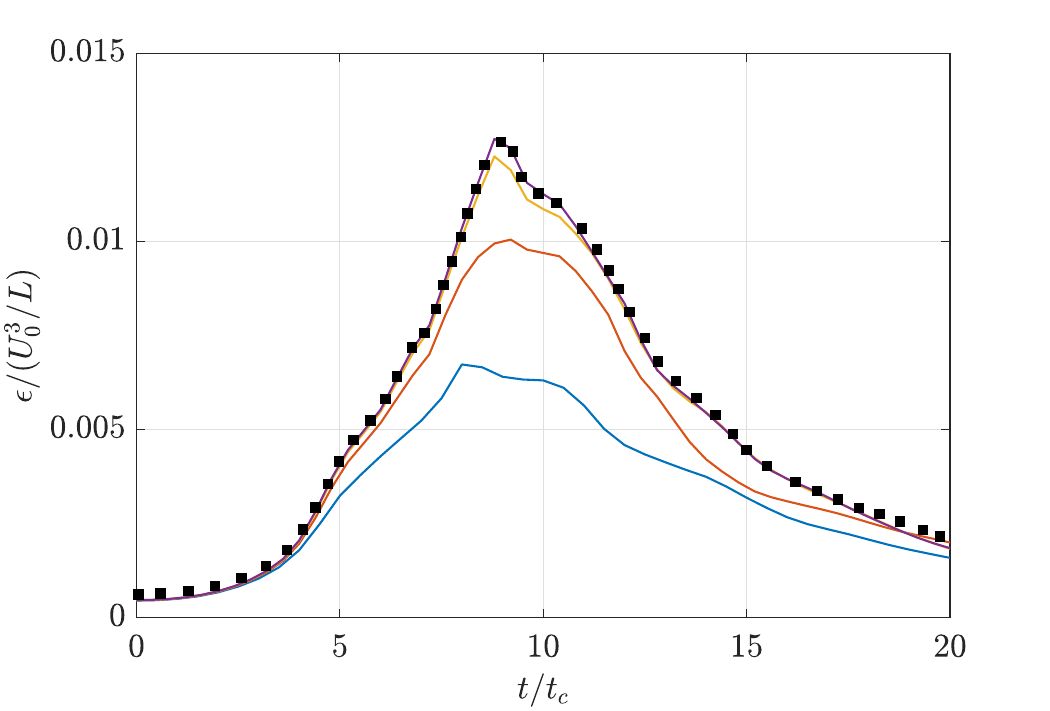}
    \caption{\label{fig:tgv3d_eps}Normalized dissipation rate.}
  \end{subfigure}
  \caption{Comparison of results for 3D Taylor-Green vortex flow, showing (a) kinetic energy and (b) dissipation. Reference data are taken from~\cite{Van_Rees2011-jz}.}
  \label{fig:tgv3d_eps_tke}
\end{figure}

The evolution of volume-averaged quantities for kinetic energy $k$ and dissipation rate, computed as $\epsilon=2\nu(\boldsymbol{S}:\boldsymbol{S})$ is illustrated  in~\cref{fig:tgv3d_eps_tke}, showing comparisons of simulation results for $N=\{128, 256, 512, 1024\}$ with the reference solution by~\citet{Van_Rees2011-jz}. The predictions of the kinetic-energy decay for the three mesh-resolutions are in good agreement with the benchmark simulations. 

Effects of the mesh resolution are more pronounced for the prediction of the dissipation rate, shown in~\cref{fig:tgv3d_eps_tke}(b). Direct comparisons with the reference data show that the coarsest mesh with $N=128$ substantially underpredicts the peak value of the dissipation rate; however, the agreement improves with increasing resolution. Only a marginal underprediction of the dissipation rate at the initial vortex decay around $8\le t/t_c\le 11$ can be seen for the simulation with intermediate mesh resolution of $N=512$, and the prediction is indistinguishable from the reference data for the finest resolution with $N=1024$. 

\begin{figure}[!htb!]
 \centering
 \begin{subfigure}[t]{0.95\columnwidth}
  \centering
  \includegraphics[width=0.32\columnwidth]{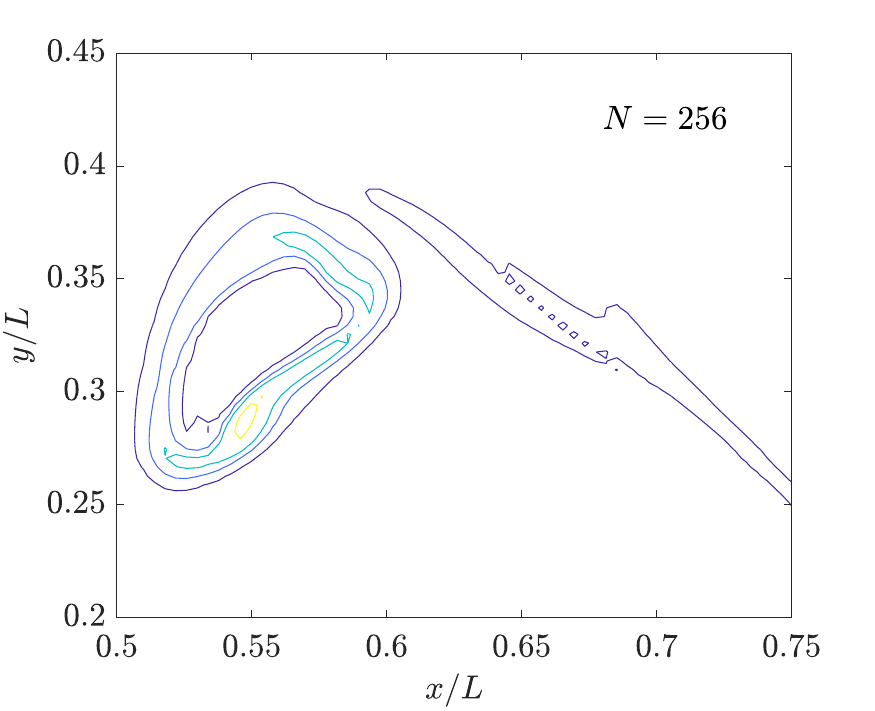}
  \includegraphics[width=0.32\columnwidth]{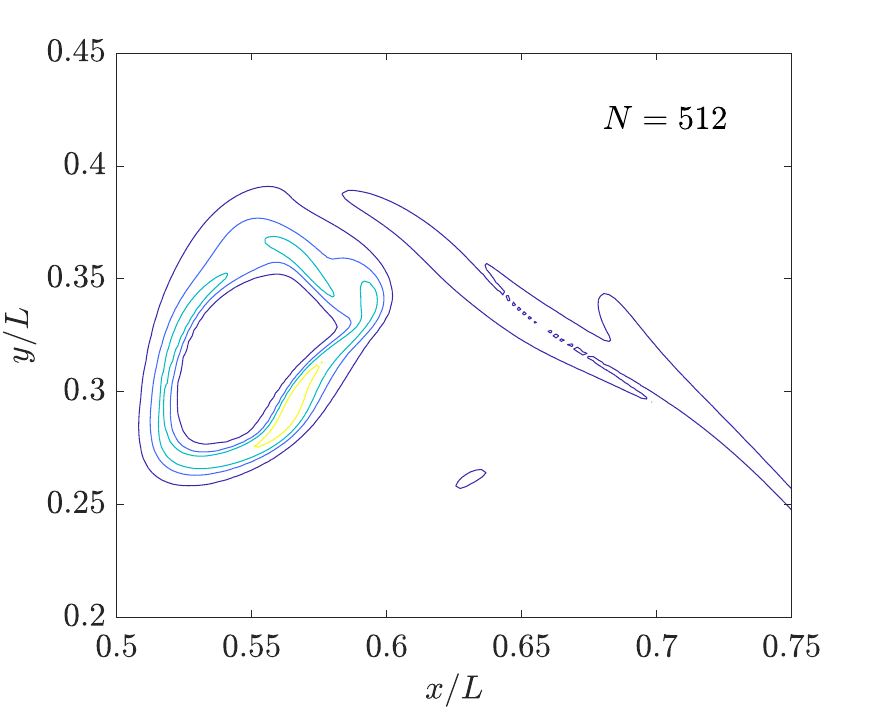}
  \includegraphics[width=0.32\columnwidth]{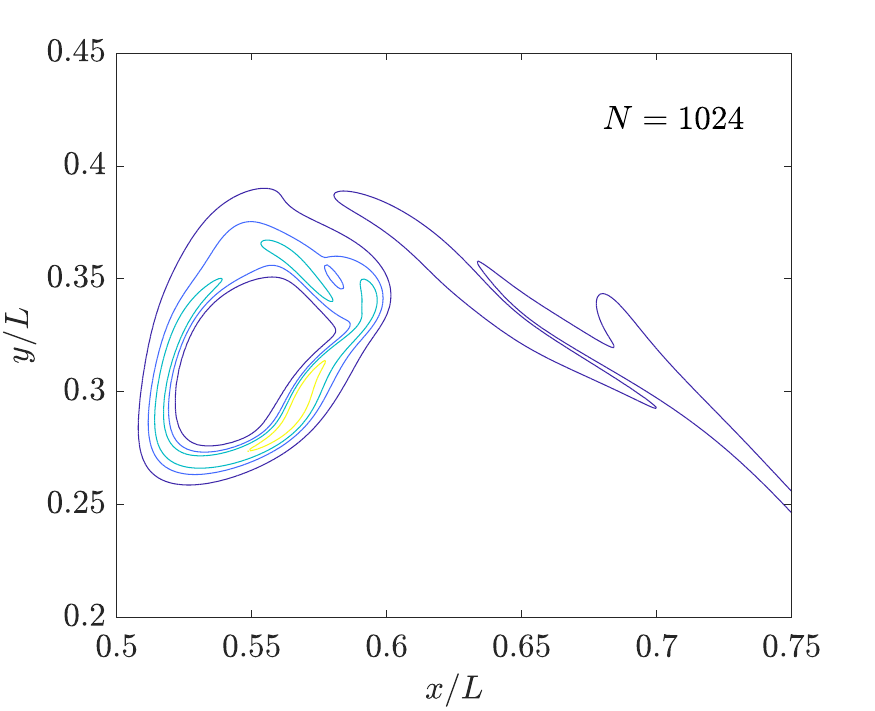}
  \caption{\label{fig:tgv3d_vorticity_t8_256_512_1024}$t/t_c=8$}
 \end{subfigure}
 \begin{subfigure}[t]{0.95\columnwidth}
  \centering
  \includegraphics[width=0.32\columnwidth]{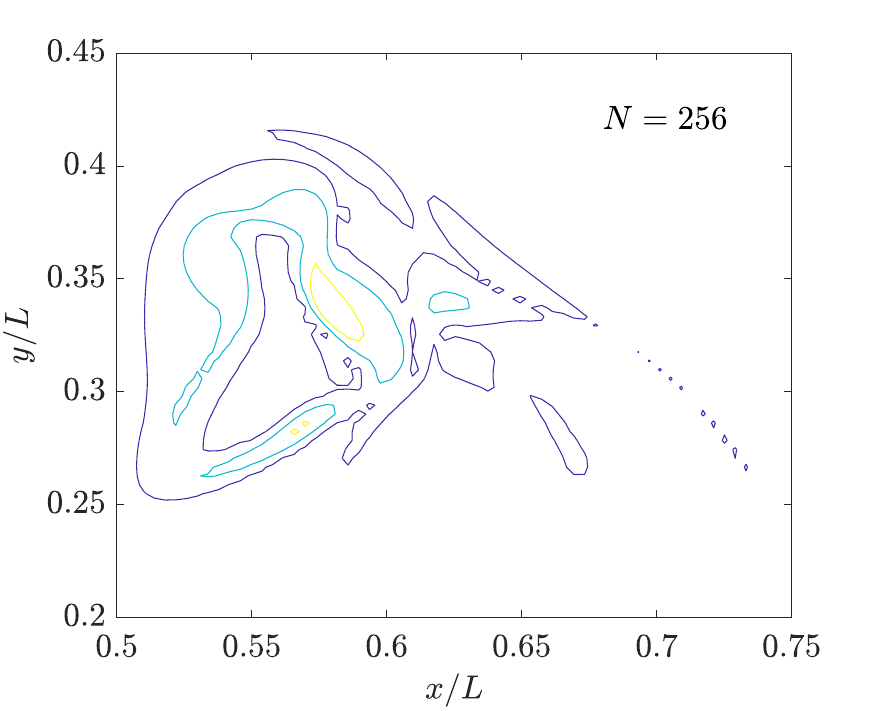}
  \includegraphics[width=0.32\columnwidth]{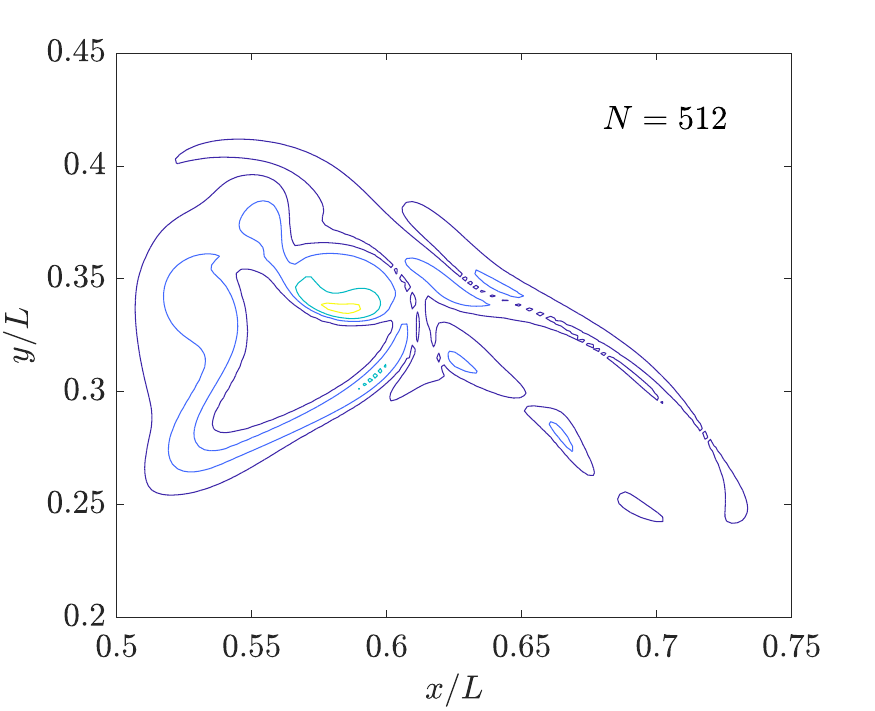}
  \includegraphics[width=0.32\columnwidth]{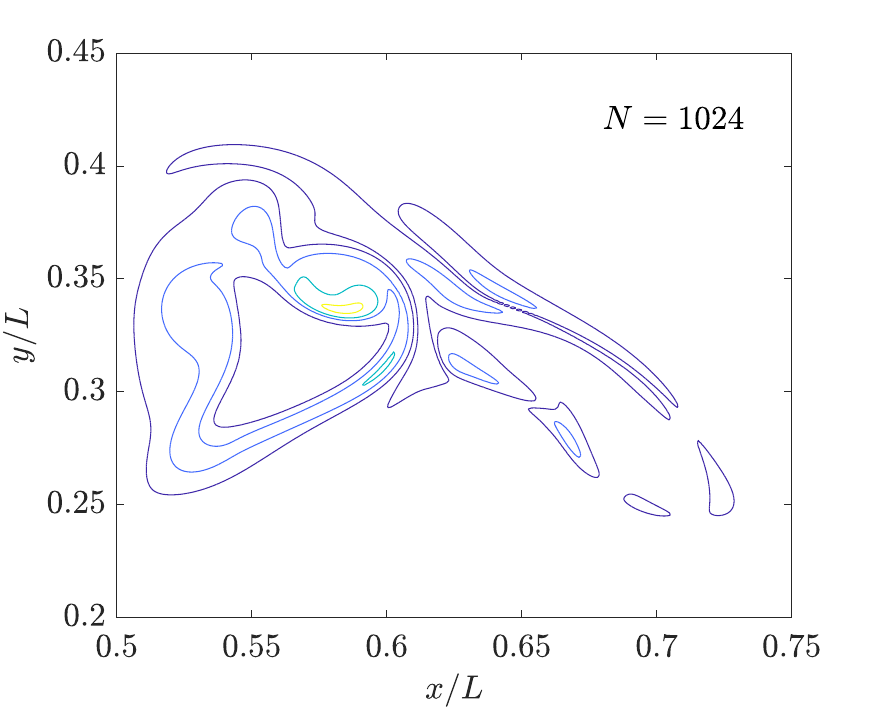}
  \caption{\label{fig:tgv3d_vorticity_t9_256_512_1024}$t/t_c=9$}
 \end{subfigure}
  \caption{Contours of vorticity magnitude, $|\omegavec|$ at the center plane $x = 0$ for (a) $t/t_c=8$ and (b) $t/t_c=9$. Results are shown for three different mesh resolutions (from left to right): $N=256, 512,$ and $1024.$ Contour values are for $|\omegavec|=\{1,5,10,20,30\}\ \text{s}^{-1}$.}
  \label{fig:tgv3d_vorticity_t8_9_256_512_1024}
\end{figure}

\Cref{fig:tgv3d_vorticity_t8_9_256_512_1024} shows instantaneous fields of the vorticity magnitude for $t/t_c=8$ and $t/t_c=9$. The large-scale vortical structures are well captured with all mesh resolutions. However, differences in capturing the smaller structures and regions with steeper gradients are not resolved by the coarsest mesh. These discrepancies become more prominent as the flow evolves and numerical errors accumulate. This can be seen in~\cref{fig:tgv3d_vorticity_t9_256_512_1024} for the simulation with $N=256$, showing the dispersion of the vortex core and excessive dissipation of the small-scale structures.

\begin{figure}[!htb!]
 \centering
 \begin{subfigure}[t]{0.49\columnwidth}
  \centering
  \includegraphics[clip,trim=0mm 0mm 12mm 4mm,width=\columnwidth]{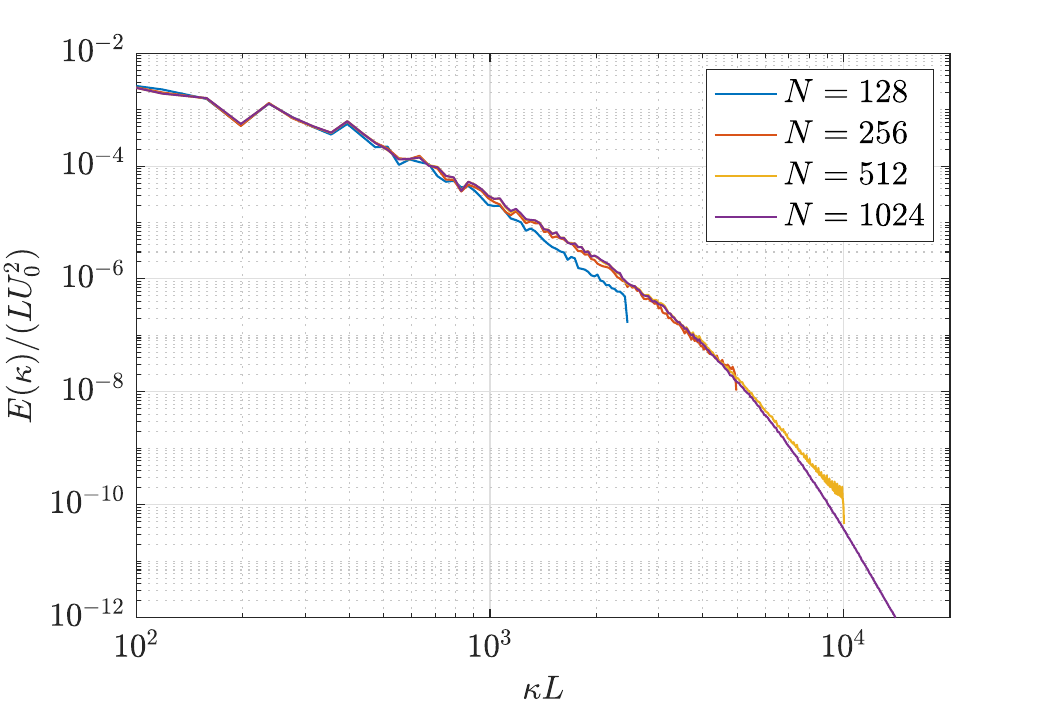}
  \caption{\label{fig:tgv3d_energy_spectrum_t8}$t/t_c=8$.}
 \end{subfigure}
 \hfill
 \begin{subfigure}[t]{0.49\columnwidth}
  \centering
  \includegraphics[clip,trim=0mm 0mm 12mm 4mm,width=\columnwidth]{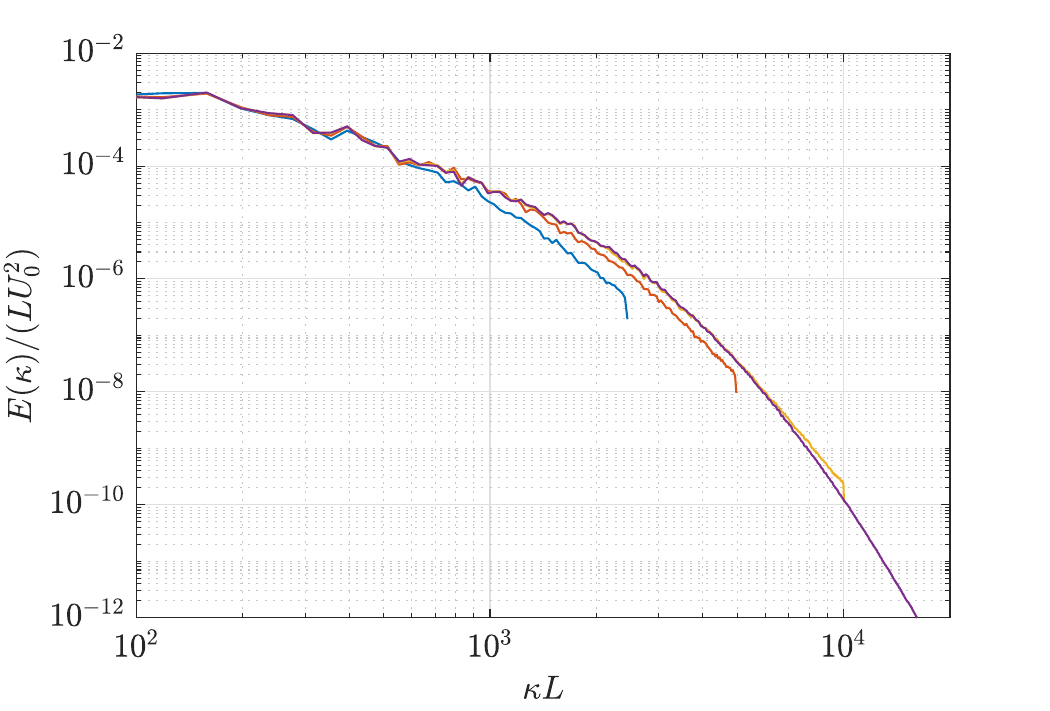}
  \caption{\label{fig:tgv3d_energy_spectrum_t9}$t/t_c=9$.}
 \end{subfigure}
 \caption{Energy spectrum for 3D Taylor-Green vortex flow at (a) $t/t_c=8$ and (b) $t/t_c=9$.}
  \label{fig:tgv3d_energy_spectrum_t8_t9}
\end{figure}

The effect of the mesh resolution is further examined by computing the energy spectra, shown in~\cref{fig:tgv3d_energy_spectrum_t8_t9} for two time instances. Apart from the solution with the coarsest mesh ($N=128$), the energy spectra at low wave numbers are identical. Differences in the spectra at higher wave numbers become apparent as smaller scales are populated by viscous-dissipative effects. The energy spectra for the simulations on the two finest meshes are comparable, indicating adequate mesh resolution.  

\begin{figure}[!htb!]
  \centering
  \includegraphics[width=0.48\textwidth]{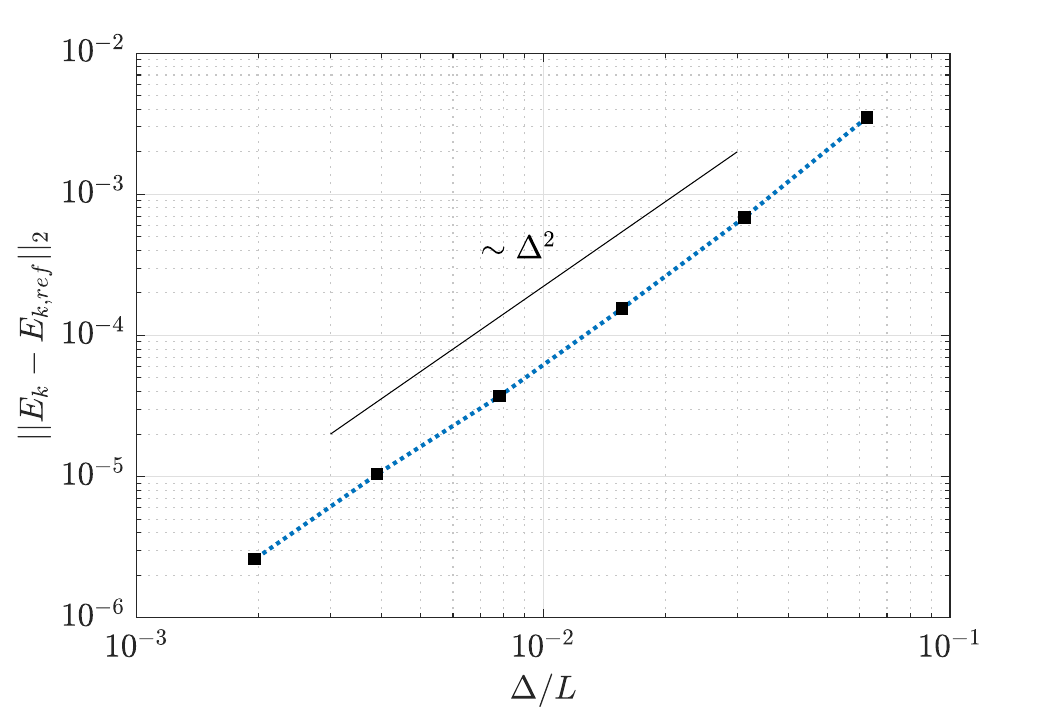}
  \caption{\label{FIG_3DTGV_TKE}Convergence for turbulent kinetic energy as a function of mesh sizes $N$.}
\end{figure}

\Cref{FIG_3DTGV_TKE} shows the rate of convergence for the kinetic energy  as a function of the mesh size $N$. The reference solution is obtained from Richardson extrapolation~\citep{Meana-Fernandez2019-gs} based on simulation results from the two finest mesh resolutions, which is:
\begin{equation}
    k_{\text{ref}} = k_N + \f{k_N-k_{N/2}} {2^2 - 1}\;,
\end{equation}
where $N=1024.$
A second-order convergence rate is observed, which is consistent with the results shown in~\cref{SSEC_2DTGV} and the discretization order of the algorithm.

\subsection{\label{SSEC_HIT}Homogeneous isotropic turbulence}
The third test case we consider is decaying homogeneous isotropic turbulence. This configuration is selected to examine the simulation accuracy in predicting the velocity spectra and the dissipative behavior. The cubic computational domain with side length $L=10.24\,\text{m}$ is discretized with $N=2048$ grid points in each direction, resulting in a homogeneous grid spacing of $\Delta = 5\times10^{-3}\,\text{m}.$ The kinematic viscosity is $\nu=10^{-3}\,\text{m}^2/\text{s}.$ In this simulation, the velocity field is initialized by a model spectrum~\cite{POPE_BOOK2000,FOX_BOOK2003}:
\begin{equation}
 \label{EQ_ENERGY_SEC}
    E(\kappa)=C_E\epsilon^{2/3}\kappa^{-5/3}f_l(\kappa l)f_\eta(\kappa \eta)
\end{equation}
with the large- and small-scale cut-off functions given as:
\begin{subequations}
 \label{EQ_SPEC_FL_FETA}
 \begin{align}
 \label{EQ_SPEC_FL}
 f_l(\kappa l) &= \left(\kappa l \left[(\kappa l)^2+c_l\right]^{-1/2}\right)^{11/3}\;,\\
 \label{EQ_SPEC_FETA}
 f_\eta(\kappa \eta) &= \exp\left\{-\beta[(\kappa\eta)^4+c^4_\eta]^{1/4}-\beta c_\eta\right\}\;,
\end{align}
\end{subequations}
and the model coefficients taken as~\cite{POPE_BOOK2000} $C_E=1.5, c_l = 6.78, \beta=5.2,$ and $c_\eta=0.4$. 

The simulation is initialized with the velocity spectrum from~\cref{EQ_ENERGY_SEC} using values for the integral length scale $l=2\,\text{m}$, Kolmogorov length scale $\eta=2\times 10^{-3}\,\text{m}$, and dissipation rate $\epsilon= \nu^3/\eta^4$. With these initial conditions, the simulation is advanced 
until the turbulence is equilibrated, resulting in the conditions $Re_\lambda=309$, $k_0=24.42\,\text{m}^2/\text{s}^2$, $\lambda_0/L=7.49\times10^{-3}$ and $l_0/L=2.84\times10^{-1}$, with $\lambda$ and $l$ being the Taylor length scale and integral length scale, respectively.  After this, the time is reset and the simulation is continued for $20\tau_l$, where the eddy-turn-over time is computed as $\tau_l=20 l_0^2/(3 Re^2_\lambda\nu)$.

\begin{figure}[!htb!]
  \centering
  \begin{subfigure}{0.32\columnwidth}
  {\includegraphics[clip,trim=4mm 1mm 10mm 0mm,width=\columnwidth]{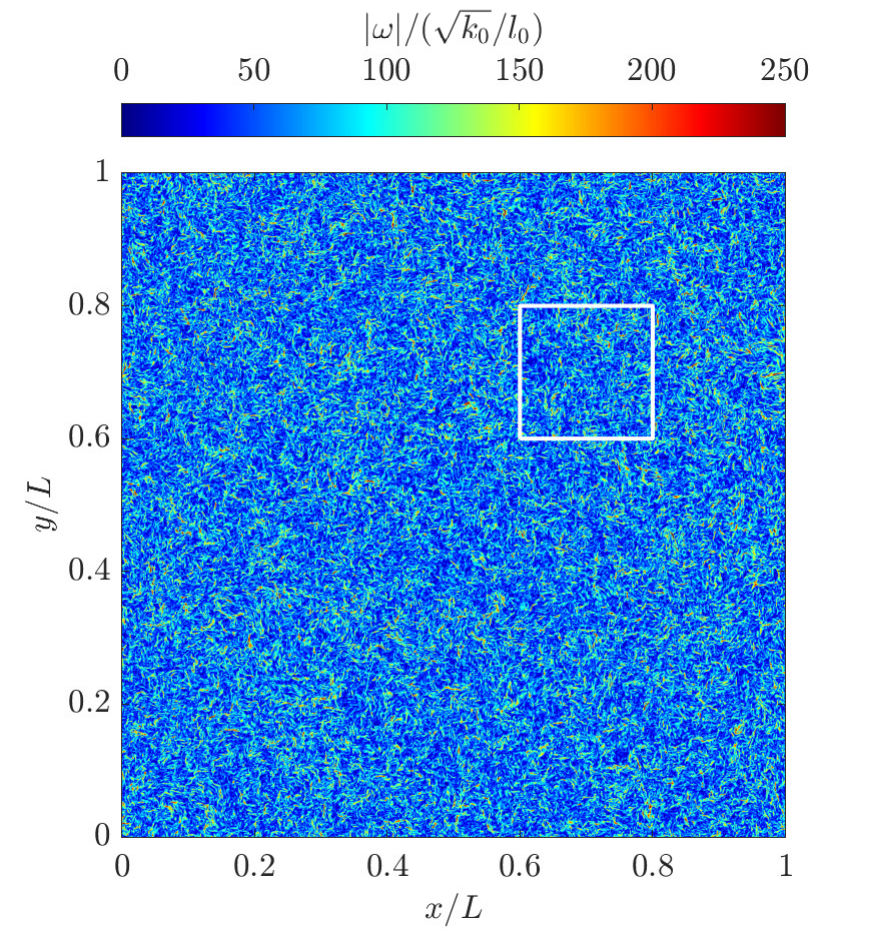}}
  \caption{$t/\tau_l=0$\label{fig:hit_vorticity_0}}
 \end{subfigure}
 \hfill
 \begin{subfigure}{0.32\columnwidth}
  {\includegraphics[clip,trim=4mm 1mm 10mm 0mm,width=\columnwidth]{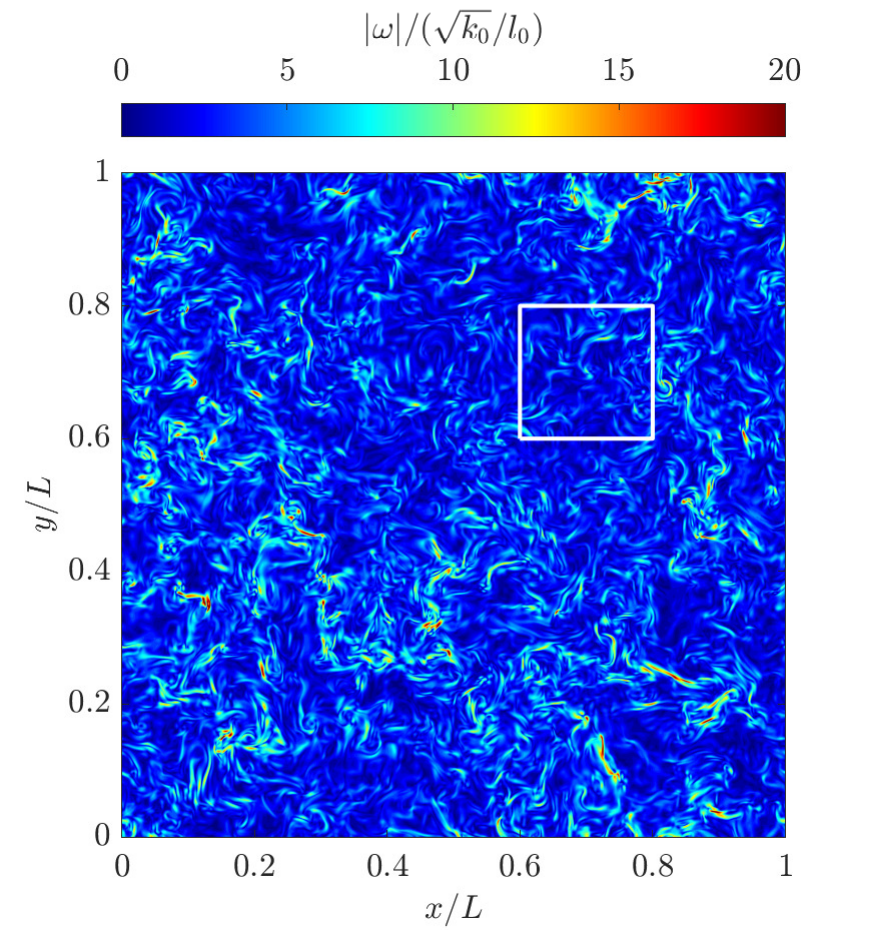}}
  \caption{$t/\tau_l=10$\label{fig:hit_vorticity_3}}
 \end{subfigure}
 \hfill
 \begin{subfigure}{0.32\columnwidth}
  {\includegraphics[clip,trim=4mm 1mm 10mm 0mm,width=\columnwidth]{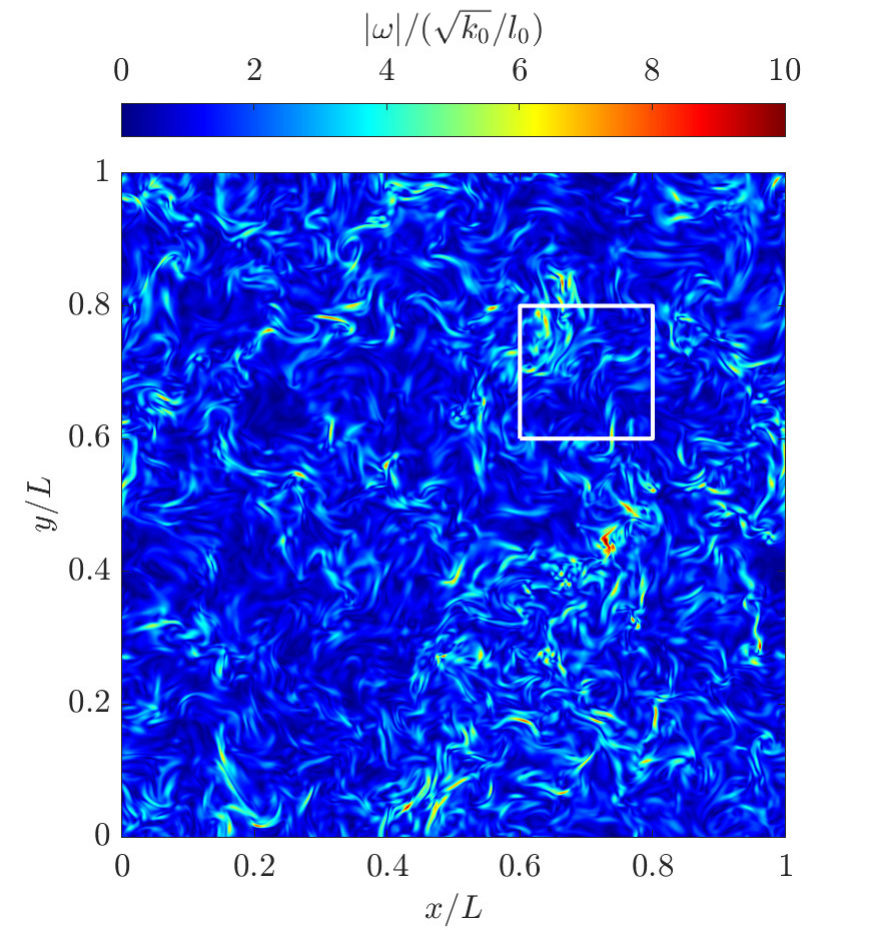}}
  \caption{$t/\tau_l=20$\label{fig:hit_vorticity_6}}
 \end{subfigure}
 \begin{subfigure}{0.32\columnwidth}
  {\includegraphics[clip,trim=1mm 1mm 10mm 0mm,width=\columnwidth]{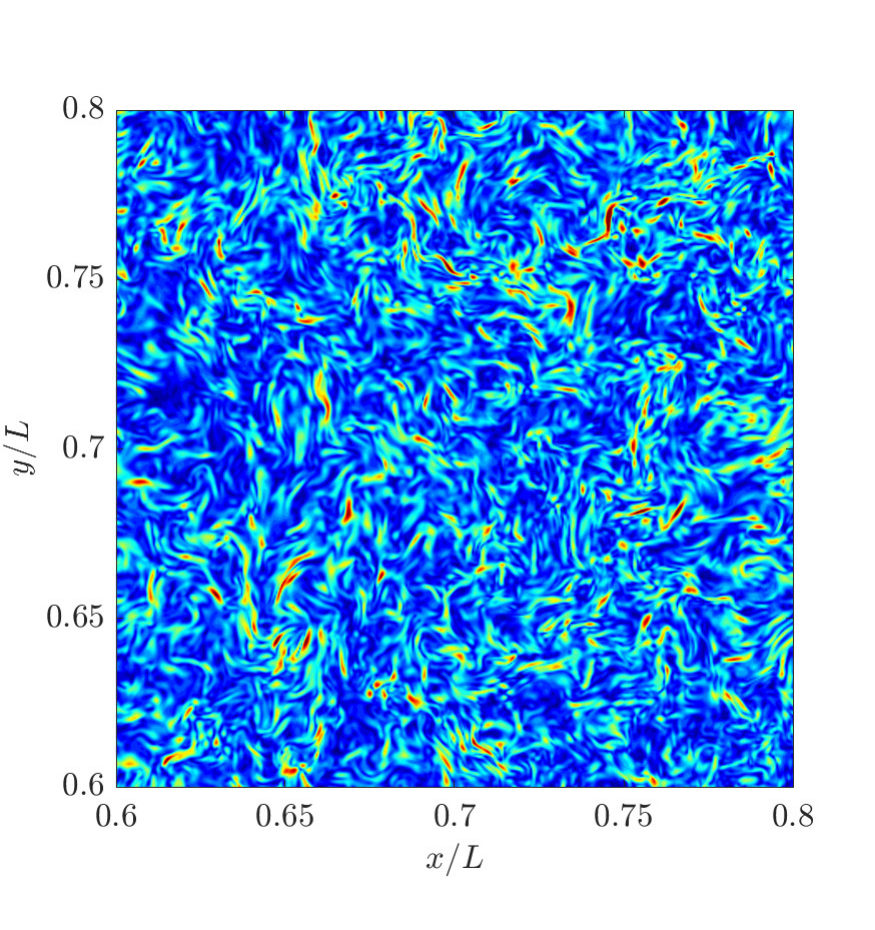}}
  \caption{$t/\tau_l=0$, $x/L\times y/L\in[0.6, 0.8]^2$\label{fig:hit_vorticity_0_zoom}}
 \end{subfigure}
 \hfill
 \begin{subfigure}{0.32\columnwidth}
  {\includegraphics[clip,trim=1mm 1mm 10mm 0mm,width=\columnwidth]{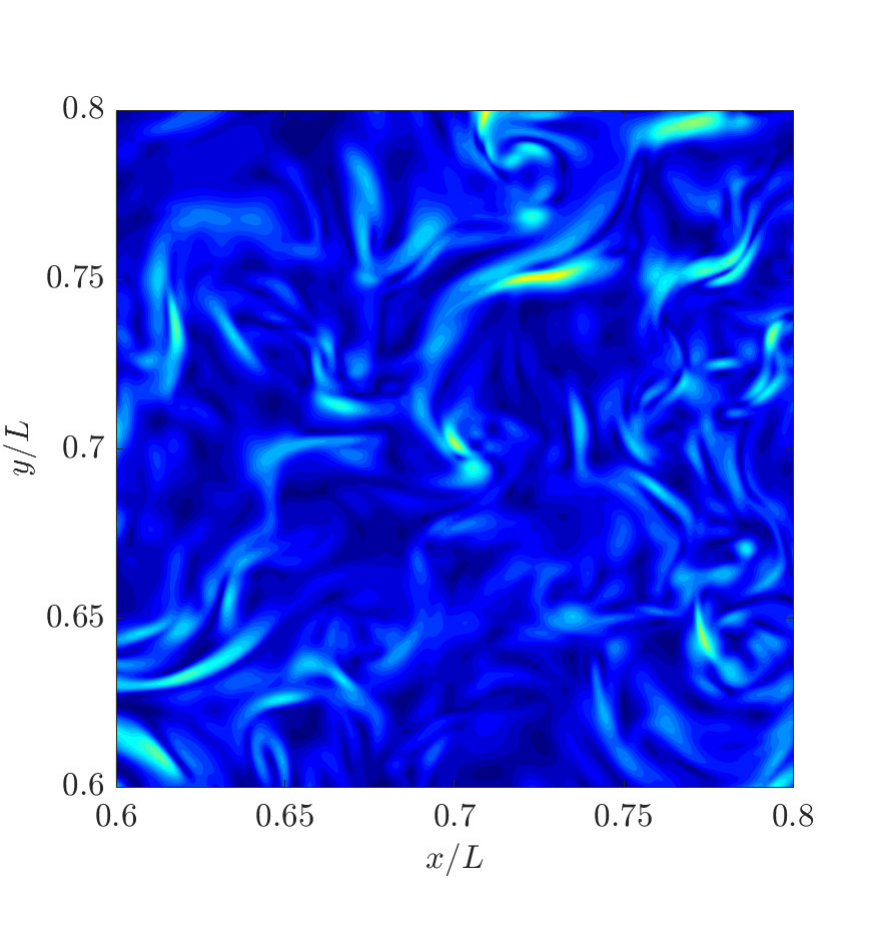}}
  \caption{$t/\tau_l=10$, $x/L\times y/L\in[0.6, 0.8]^2$\label{fig:hit_vorticity_3_zoom}}
 \end{subfigure}
 \hfill
 \begin{subfigure}{0.32\columnwidth}
  {\includegraphics[clip,trim=1mm 1mm 10mm 0mm,width=\columnwidth]{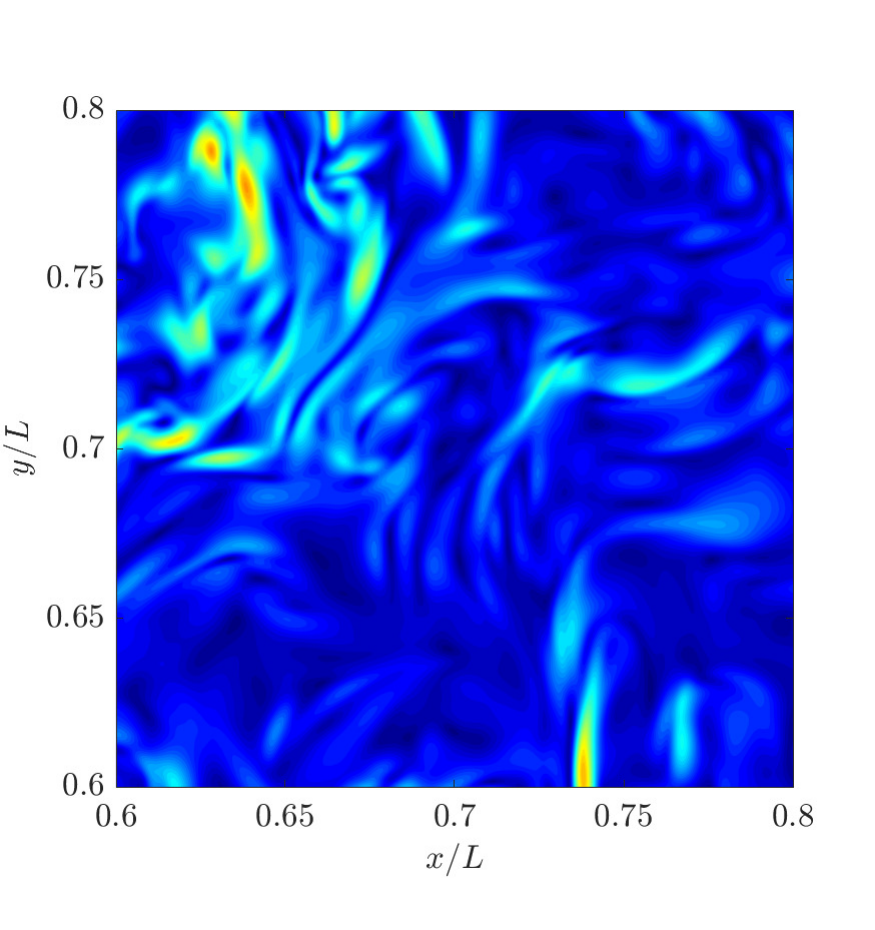}}
  \caption{$t/\tau_l=20$, $x/L\times y/L\in[0.6, 0.8]^2$\label{fig:hit_vorticity_6_zoom}}
 \end{subfigure}
  \caption{Instantaneous flow-field results of vorticity magnitude for homogeneous isotropic turbulence at $z/L=0.5$ for (a) $t/\tau_l=0$, (b)  $t/\tau_l=10$ and (c) $t/\tau_l=20$. Bottom row shows zoom of vorticity in white box (a, d), (b, e), and (c, f).}
  \label{fig:hit_vel_vorticity}
\end{figure}
\begin{figure}[!htb!]
  \centering
  \begin{subfigure}[t]{0.49\columnwidth}
  \centering
  \includegraphics[width=\columnwidth]{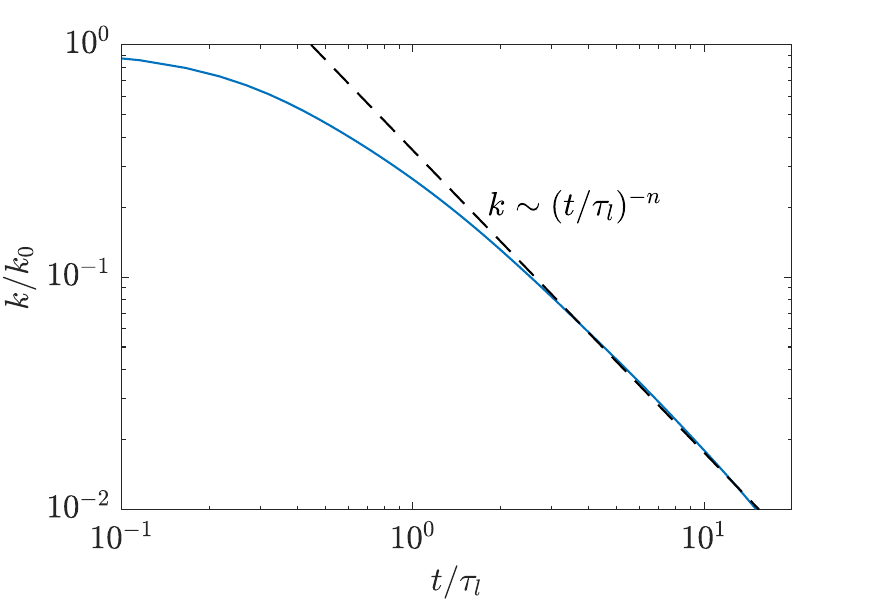}
  \caption{\label{fig:hit_tke}Turbulent kinetic energy.}
 \end{subfigure}
 \begin{subfigure}[t]{0.49\columnwidth}
  \centering
  \includegraphics[width=\columnwidth]{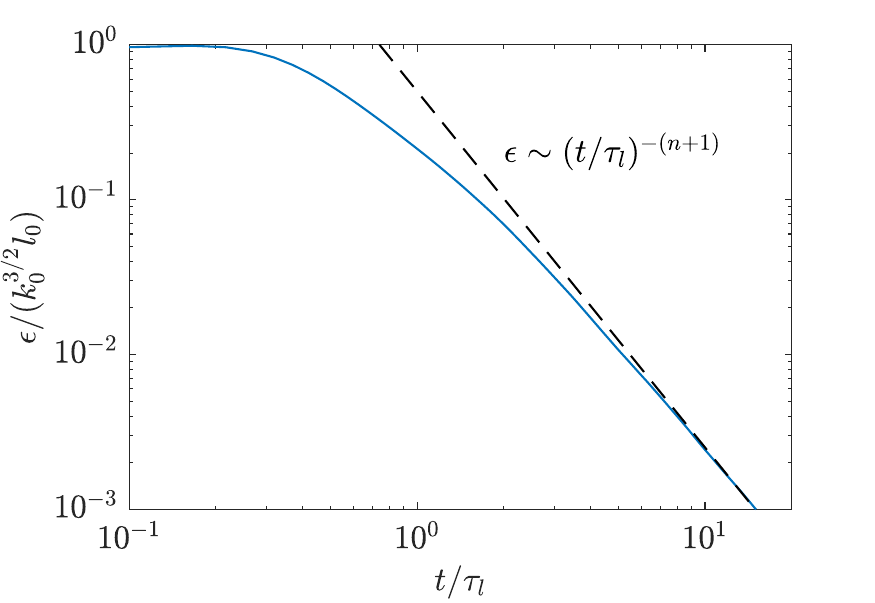}
  \caption{\label{fig:hit_epsilon}Dissipation rate.}
 \end{subfigure}
 \caption{\label{FIG_HIT_STATS_K_EPS}Temporal evolution of (a) the turbulent kinetic energy and (b) dissipation rate. Results are compared with analytical power laws for decaying homogeneous isotropic turbulence (dashed lines).}
\end{figure}

Instantaneous simulation results for the vorticity magnitude are presented in~\cref{fig:hit_vel_vorticity} for three different time-instances, corresponding to $t/\tau_l=0, 10,$ and $20$. In the absence of intrinsic turbulence production, it can be seen that the vorticity decays in time, resulting in an increase in the spatial coherence. 
To provide a quantitative assessment of the simulation results, we present profiles for the temporal evolution of the turbulent kinetic energy and dissipation rate in~\cref{FIG_HIT_STATS_K_EPS}. The decay of the turbulent kinetic energy follows a power-law with $k/k_0\sim\left(t/\tau_l\right)^{-n}$. Since the dissipation rate is given by the equation $\epsilon=dk/dt$, the decay of the dissipation takes the expression $\epsilon/\epsilon_0\sim\left(t/\tau_l\right)^{-(n+1)}$. For $t/\tau_l>2$ the coefficient of decay is found to be $n=1.3$, which is in agreement with literature~\cite{POPE_BOOK2000}.

\begin{figure}[!htb!]
  \centering
  \includegraphics[width=0.5\textwidth]{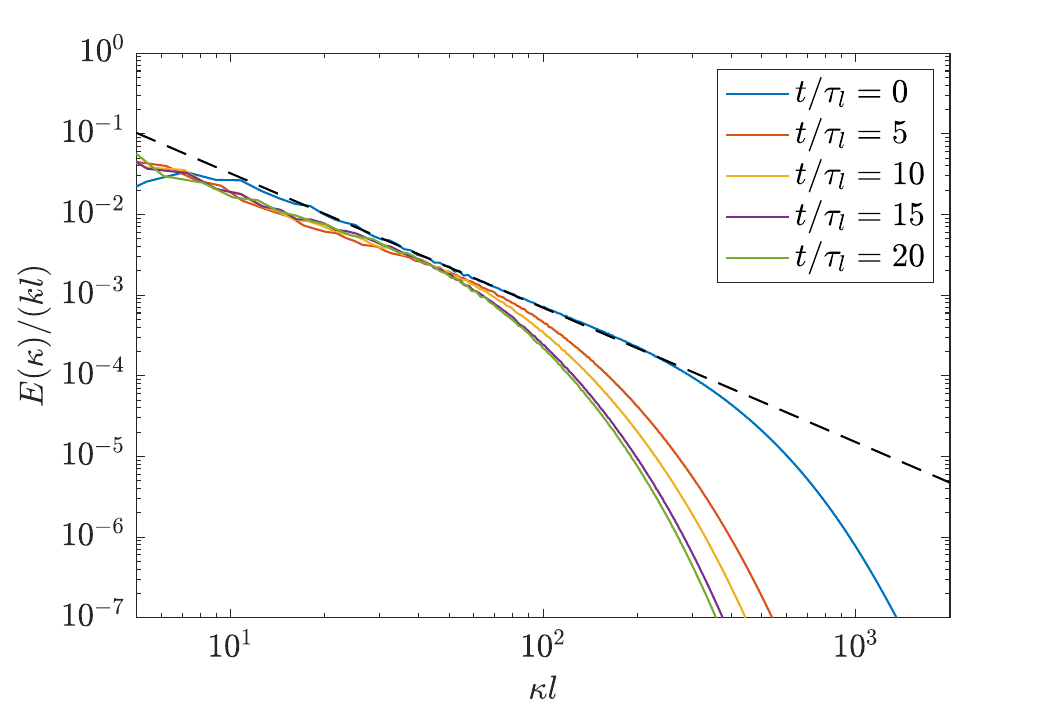}
  \caption{Turbulent energy spectrum computed from decaying homogeneous isotropic turbulence for different time instances. Spectra are normalized with local quantities for $l(t)$ and $k(t)$.}
  \label{fig:hit_spectrum}
\end{figure}

\Cref{fig:hit_spectrum} compares the kinetic energy spectra for different time instances during the simulation. The spectra are normalized by the instantaneous integral length $l(t)$ and the turbulent kinetic energy $k(t)$. The results show evidence of an inertial subrange that follows the expected $-5/3$ decay rate (indicated by the dashed line), followed by the viscous-dissipative range. Because of the initialization of the flow, the energy-containing range is shifted towards smaller wave numbers and is not fully represented in this simulation. 

\subsection{\label{SSEC_JET}Planar turbulent jet flow}
The last test case that we considered in this work is concerned with simulating a planar turbulent jet. This configuration was studied experimentally and numerically by~\citet{Watanabe2012-sc,Watanabe2013-yz,Watanabe2014-wh}, and provides reference data for comparisons.  In this configuration, a passive scalar is ejected from a planar horizontal slot of width $D=2\,\text{mm}$,  at a mean velocity of $U_J=1.29\,\text{m/s}$ into a coflow of air at a velocity of  $U_A/U_J=0.056$. The jet-exit Reynolds number is $Re_J=DU_J/\nu=2300$. The computational domain is $60\,D\times 50\,D\times 7.5\,D$ in streamwise, spanwise, and lateral direction, respectively. Free-slip boundary conditions are applied along spanwise directions and periodic boundary conditions are applied along the later direction of the simulation domain. Convective outflow conditions are prescribed at the exit of the domain. The domain is discretized with a uniform mesh using $1280\times 1024\times 128$ grid points along streamwise, spanwise, and lateral directions. The jet velocity inflow conditions are prescribed by a turbulent profile to replicate the experimental conditions. Random perturbations with root-mean-square velocity fluctuations obtained from the measurements are imposed on the mean inflow profile. Simulations are performed for 10 flow-through times and statistics are collected over the last five flow-through times.

\begin{figure}[!htb!]
  \centering
  \begin{subfigure}{0.49\columnwidth}
  \centering
  {\includegraphics[clip,trim=0mm 0mm 10mm 0mm,width=\columnwidth]{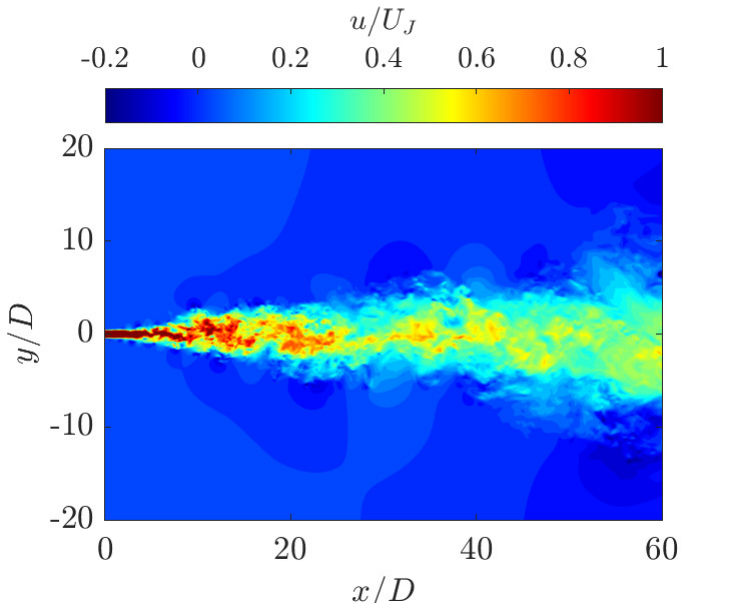}}
  \caption{\label{fig:jet_contour_u}Axial velocity}
 \end{subfigure}
 \hfill
 \begin{subfigure}{0.49\columnwidth}
  \centering
  {\includegraphics[clip,trim=0mm 0mm 10mm 0mm,width=\columnwidth]{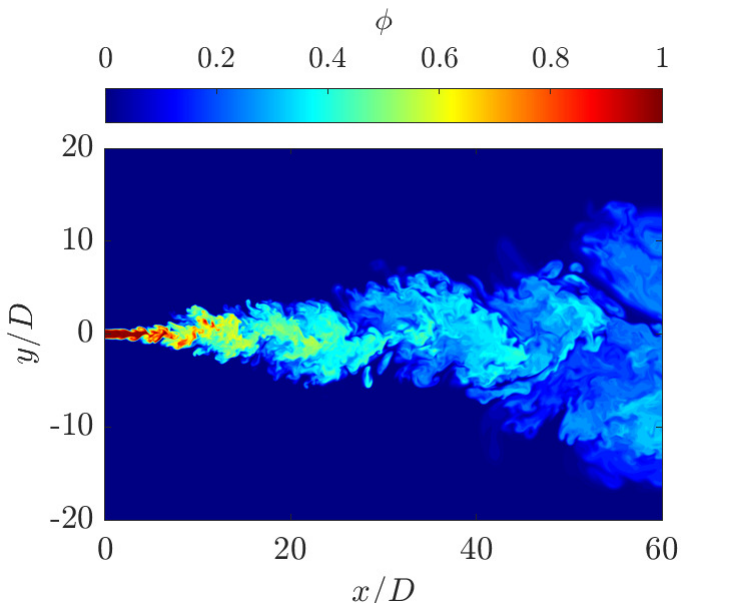}}
  \caption{\label{fig:jet_contour_z}Passive scalar}
 \end{subfigure}
 \begin{subfigure}{0.49\columnwidth}
  \centering
  {\includegraphics[clip,trim=0mm 0mm 10mm 0mm,width=\columnwidth]{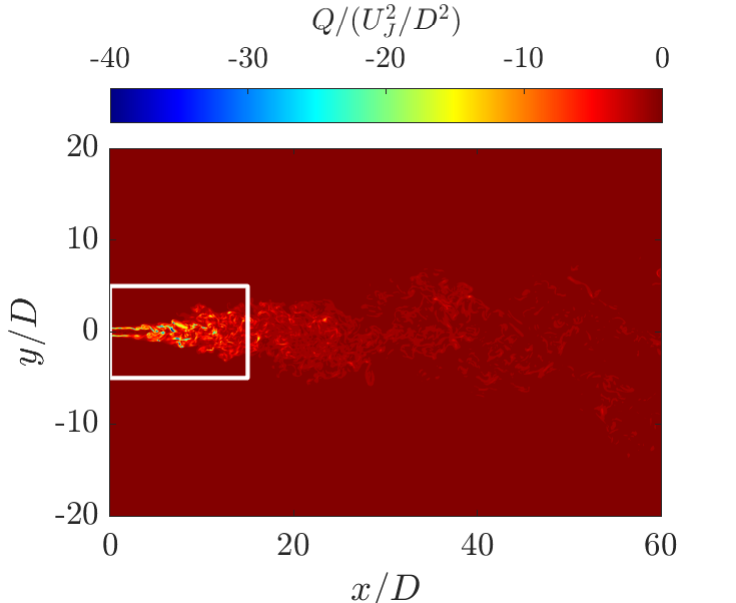}}
  \caption{\label{fig:jet_contour_q_crit}$Q$-criterion}
 \end{subfigure}
 \hfill
 \begin{subfigure}{0.49\columnwidth}
  \centering
  {\includegraphics[clip,trim=0mm 0mm 10mm 0mm,width=\columnwidth]{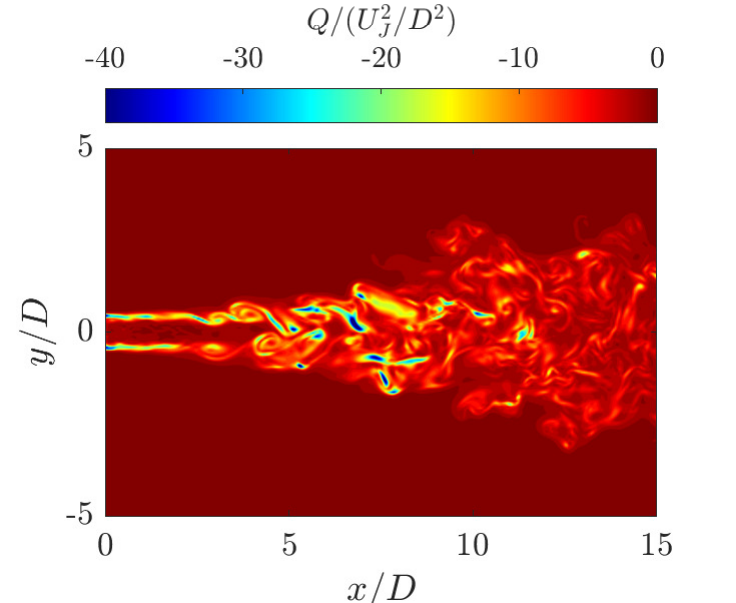}}
  \caption{\label{fig:jet_contour_q_crit_zoom_in}Zoom of $Q$-criterion}
 \end{subfigure}
  \caption{Instantaneous flow field results for (a) axial velocity, (b) passive scalar $\phi$, (c) $Q$-criterion and (d) zoom in nozzle-near region.}
  \label{fig:jet_contour_u_z_phi_q_qzoom}
\end{figure}

\Cref{fig:jet_contour_u_z_phi_q_qzoom} shows instantaneous flow-field results for axial velocity, passive scalar and $Q$-criterion along the center plane of the domain. From these results, the formation of a potential core is visible that closes after $5D$ by the developing shear-layer instability and is followed by the transition to a fully developed turbulent flow. The entrainment of air from the coflow into the jet results in the mixing and spreading of the jet along the downstream direction. The jet-spreading is accompanied by the formation of large-scale vortical structures that break up as the scalar is advected in downstream direction.

\begin{figure}[!htb!]
  \centering
  \begin{subfigure}{0.49\columnwidth}
  \centering
  \includegraphics[width=\columnwidth]{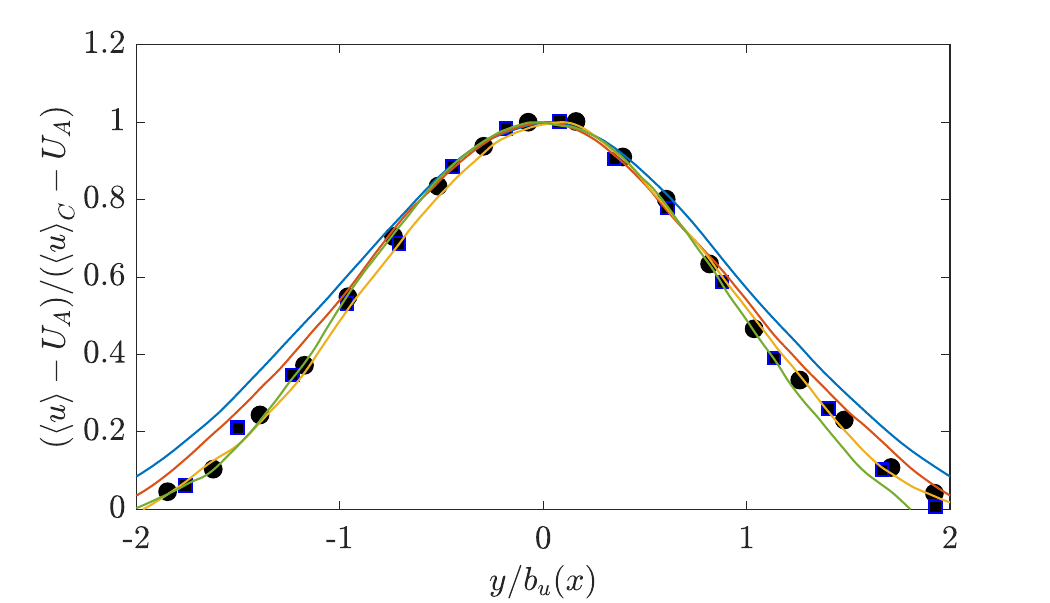}
  \caption{\label{fig:jet_similarity_u}Axial velocity}
 \end{subfigure}
 \begin{subfigure}{0.49\columnwidth}
  \centering
  \includegraphics[width=\columnwidth]{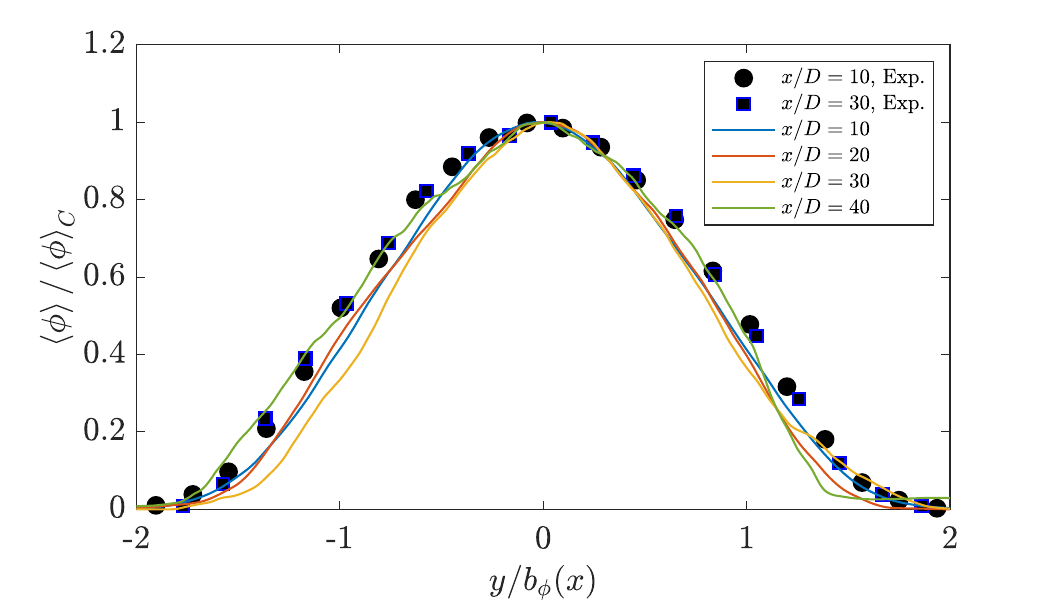}
  \caption{\label{fig:jet_similarity_z}Passive scalar}
 \end{subfigure}
  \caption{Comparison of self-similar profiles for (a) axial velocity and (b) passive scalar between simulations and experiments~\citep{Watanabe2012-sc}.}
  \label{fig:jet_similarity}
\end{figure}

\begin{figure}[!htb!]
  \centering
  \begin{subfigure}{0.49\columnwidth}
  \centering
  \includegraphics[width=\columnwidth]{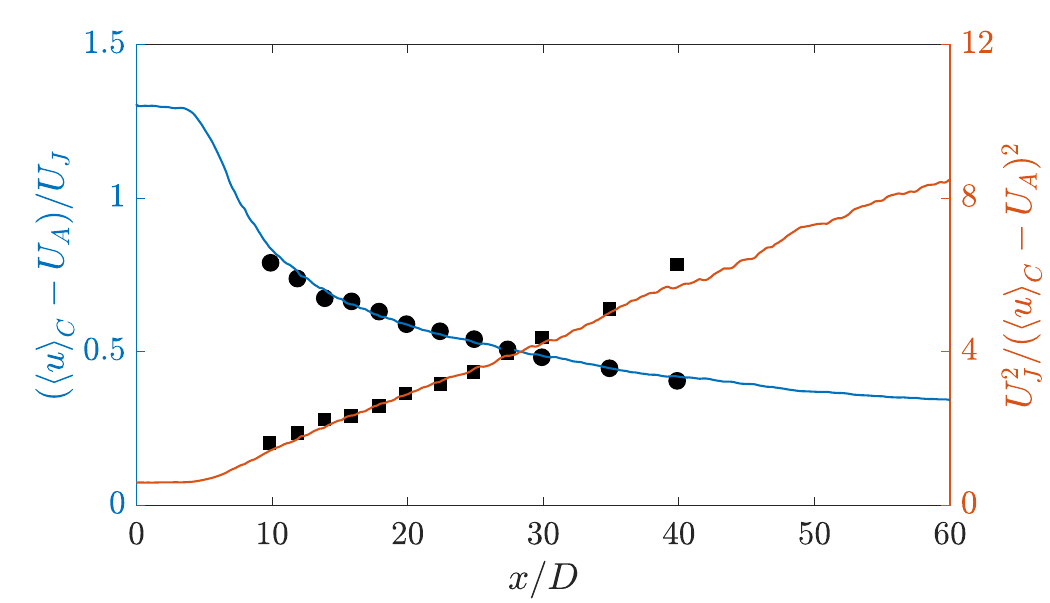}
  \caption{\label{fig:jet_center_u}Decay of axial velocity}
 \end{subfigure}
 \begin{subfigure}{0.49\columnwidth}
  \centering
  \includegraphics[width=\columnwidth]{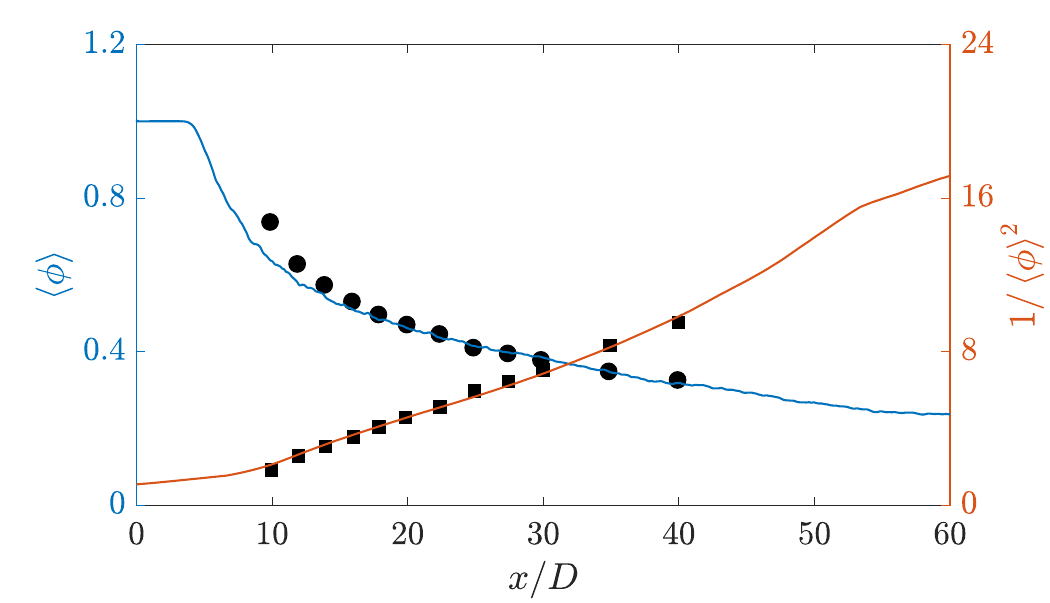}
  \caption{\label{fig:jet_center_z}Decay of passive scalar}
 \end{subfigure}
  \caption{Streamwise variation in (a) mean streamwise velocity, and (b) mean passive scalar mass fraction on jet centerline. Simulation results are compared with experiments~\citep{Watanabe2012-sc}.}
  \label{fig:jet_center}
\end{figure}

\Cref{fig:jet_similarity} shows mean profiles of axial velocity component and passive scalar at $x/D=\{10, 20,30, 40\}$. Both quantities are normalized by their respective centerline conditions. The cross-stream axis $y$ is normalized by the jet half width conditions~\cite{Watanabe2014-wh}:
\begin{align}
    b_u/D &= 0.079(x/D) + 0.263, \\
    b_\phi/D &= 0.118(x/D) + 0.230.
    \label{eq:half-width}
\end{align}
The solutions for the streamwise velocity and the scalar exhibit similarity, and the simulation results are in good agreement with experimental data. \Cref{fig:jet_center} shows mean profiles of the axial velocity and the scalar along the center line. Good agreement with measurements can be observed for both quantities. The rate of decay for both quantities scales with $x^{1/2}$~\cite{Watanabe2014-wh}, which matches the experiment quantitatively.

\section{\label{SEC_PERFORMANCE}Scalability analysis}
We conclude our analysis by examining the scalability of the solver. For this, we consider the 3D TGV-configuration that was discussed in~\cref{SSEC_3DTGV}. These scalability tests are performed on a TPU v3 pod with 2048 cores.

\subsection{\label{SSEC_WEAK_SCALABILITY}Weak scalability}
\begin{figure}[!htb!]
  \centering
  \includegraphics[width=0.49\textwidth]{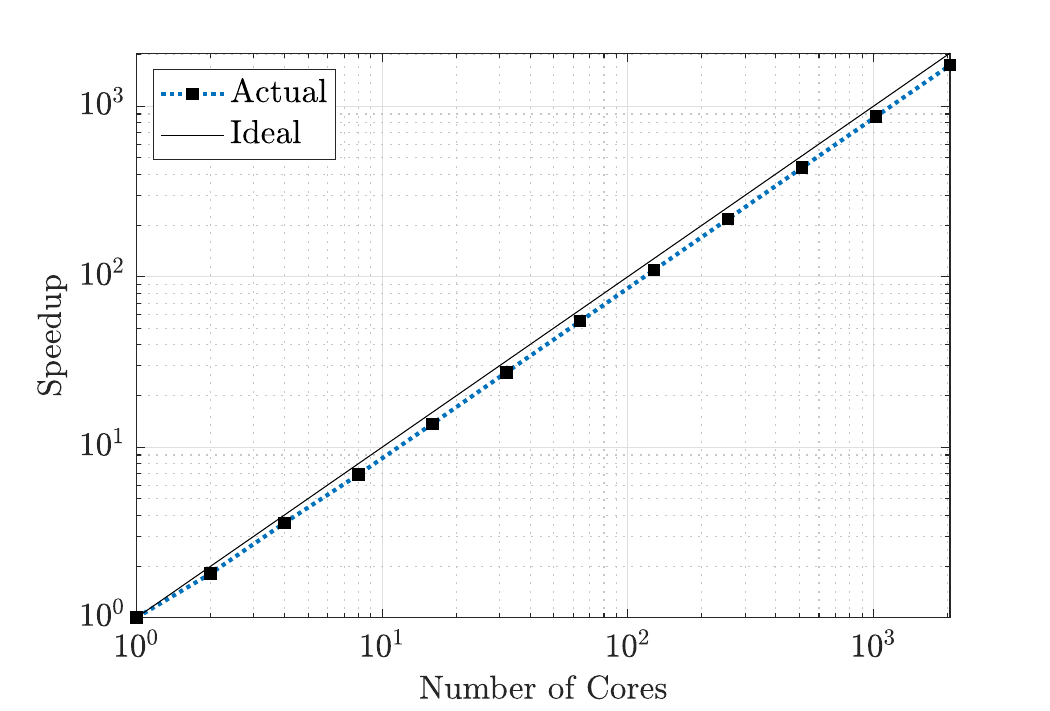}
  \caption{Weak scalability of the TPU-CFD solver using up to a full TPU v3 pod (2048 cores).}
  \label{fig:scalability_weak}
\end{figure}

For the weak scaling analysis, the computational domain per TPU core is fixed with a size of $\wh{N}_x\times \wh{N}_y\times \wh{N}_z=1024\times1024\times36$, and the total mesh size increases with the number of TPU cores. The computational domain is partitioned in such a way that at least two cores are assigned in the $x$- and $y$-direction. This is to avoid the influence of the scaling performance due to the underlying data structure. As discussed in~\cref{sssec:kernel_op}, the 3D data are represented by a python list of 2D TensorFlow tensors, with each \lstinline{tf.Tensor} representing a $x$-$y$ plane, and the list traversing the $z$-direction. Experimentally, we found that the XLA compiler is better at optimizing the TensorFlow graph partition along the list direction than in the $x$- and $y$-directions. Therefore, instead of employing a homogeneous distribution, we preferentially partition along the $z$-direction and use not more than four cores for partitioning the $x$- and $y$-direction, respectively. The partitioning and time per iteration is shown in~\cref{tab:weak_scaling}. \Cref{fig:scalability_weak} provides a graphical illustration of the speed-up, showing nearly linear speedup.

\begin{table}[!htb!]
\centering
\footnotesize
\caption{Partitions and simulation configurations for weak scalability analysis.}
\begin{tabular}{|c|c|c|c|c|c|c|c|}\hline
\multicolumn{4}{|c|}{Number of cores}&\multicolumn{3}{c|}{Partition} & $T(P_\text{tot})$\\
$P_\text{tot}$ &$P_x$ & $P_y$ & $P_z$ &  $\wh{N}_x$ & $\wh{N}_y$ & $\wh{N}_z$ & [ms]\\
     \hline\hline
     2048 & 4 & 4 & 128 & 1024 & 1024& 36 & 0.634 \\
     1024 & 2 & 4 & 128 & 1024 & 1024& 36 & 1.268 \\
     512 & 2 & 2 & 128 & 1024 & 1024& 36 & 2.529 \\
     256 & 2 & 2 & 64 & 1024 & 1024& 36 & 5.051 \\
     128 & 2 & 2 & 32 & 1024 & 1024& 36 & 10.07 \\
     64 & 2 & 2 & 16 & 1024 & 1024& 36 & 20.13 \\
     32 & 2 & 2 & 8 & 1024 & 1024& 36 & 40.19 \\
     16 & 2 & 2 & 4 & 1024 & 1024& 36 & 80.38 \\
     8 & 2 & 2 & 2 & 1024 & 1024& 36 & 160.0 \\
     4 & 2 & 2 & 1 & 1024 & 1024& 36 & 307.3 \\
     2 & 1 & 2 & 1 & 1024 & 1024& 36 & 608.5 \\
     1 & 1 & 1 & 1 & 1024 & 1024& 36 & 1203 \\
     \hline
 \end{tabular}
 \label{tab:weak_scaling}
\end{table}

\subsection{\label{SSEC_STRONG_SCALABILITY}Strong scalability}

The strong scalability analysis is conducted by considering the following two conditions: (i) the data of a single partition can fit in the memory of a TPU for the coarsest partitioning; and (ii) a single partition is not too small for the finest partitioning. Based on these conditions, two problems are selected to test the scalability: the first one takes a mesh of total size $256\times 256\times 16384$, and the second one has size $1024^3$. For both cases, the partition changes in the $z$-direction only. In the first case, one core is used in the $x$- and $y$-directions, and the partition in the $z$-direction changes from 64 to 2048 cores. In the second case, four cores are used in the $x$- and $y$-directions, respectively, and the partition in the $z$-direction changes from 4 to 128 cores. Note that the number of cores assigned along the $x$- and $y$-directions is low compared to that in the $z$-direction. This is because partitioning a \lstinline{tf.Tensor} is less efficient than partitioning a python list. Based on these observations, we recommend partitioning along the list direction as much as possible to fully utilizing the MXU-hardware architecture and optimize the TPU performance.

\begin{figure}[!htb!]
  \centering
  \begin{subfigure}{0.49\textwidth}
    \centering
    \includegraphics[width=\textwidth]{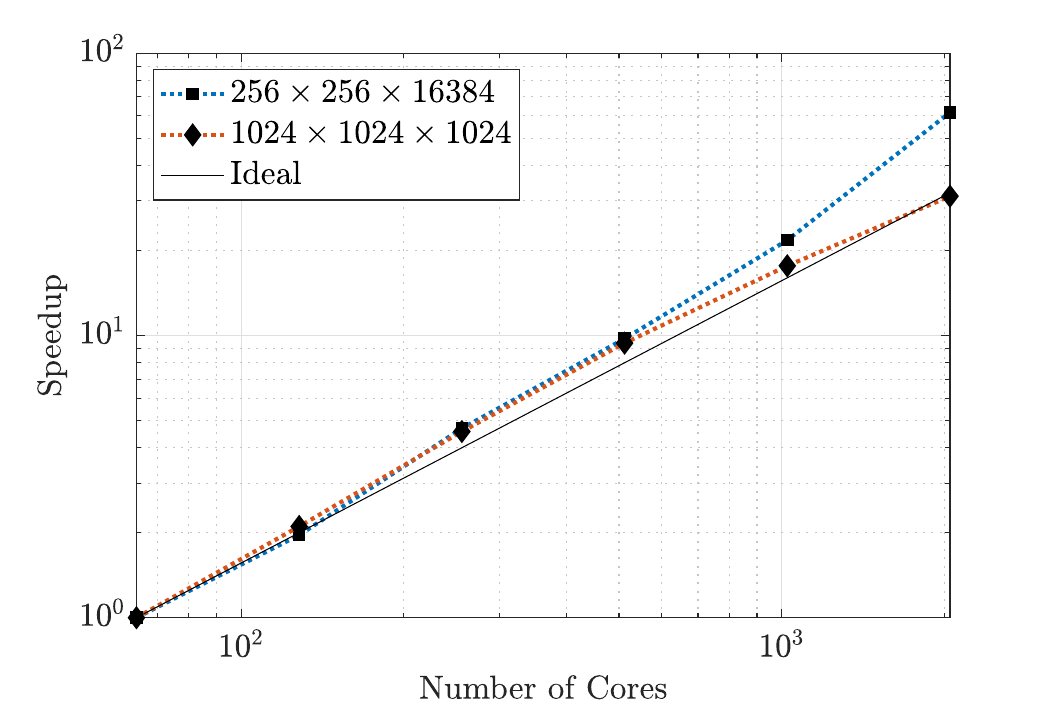}
    \caption{\label{fig:scalability_strong_1}Speedup.}
  \end{subfigure}
  \hfill
  \begin{subfigure}{0.49\textwidth}
    \centering
    \includegraphics[width=\textwidth]{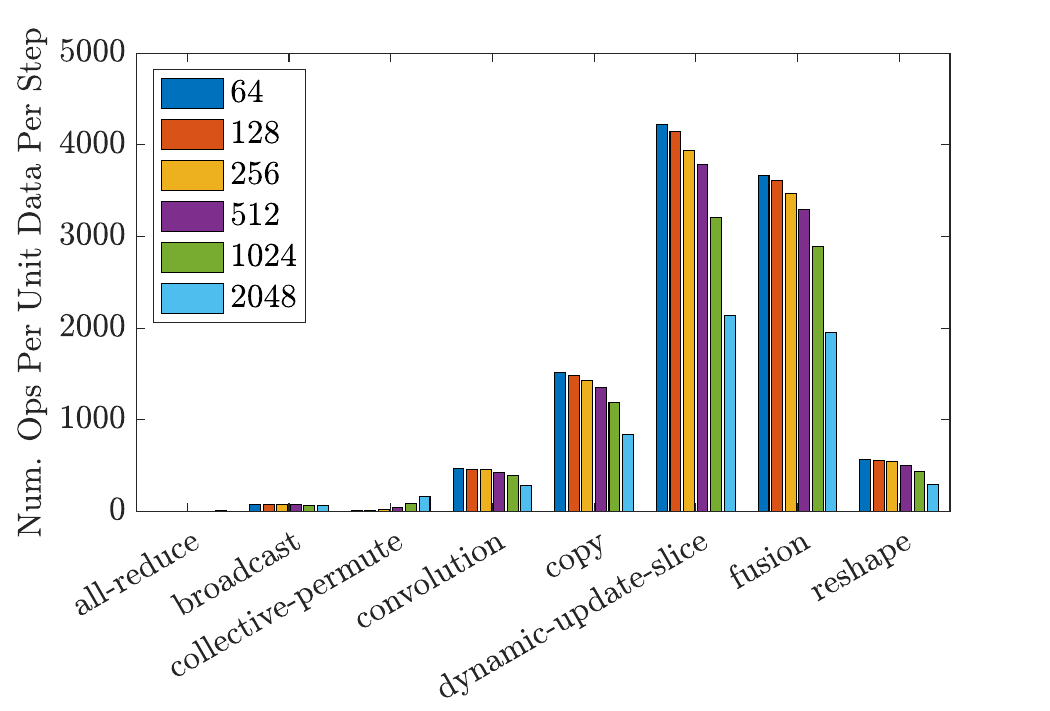}
    \caption{\label{fig:scalability_time}Operation time per time step and TPU core.}
  \end{subfigure}
  \caption{Strong scalability of TPU-CFD solver showing (a) speed-up, and (b) profiling of the $256\times 256\times 16384$ mesh points test case up to 2048 TPU-cores, corresponding to a full TPU v3 pod.}
  \label{fig:scalability_strong}
\end{figure}

Results for the speed-up are illustrated in~\cref{fig:scalability_strong_1}. The curve for the first case shows clear evidence of superlinear scaling. To examine the cause of this behavior, we profiled the code and results are presented in~\cref{fig:scalability_time}, showing wall time per time step and degree of freedom associated with the different operations that correspond to communication ({\tt{all-reduce}}, {\tt{broadcast}}, {\tt{collective-permute}}), data manipulation ({\tt{copy}}, {\tt{reshape}}), and computations ({\tt{fusion}}, {\tt{convolution}}, {\tt{dynamic-update-slice}}). Note that communication-related operations ({\tt{all-reduce}}, {\tt{broadcast}}, {\tt{collective-permute}}) are associated with exchanges of boundary conditions and global numerical operations such as the mean computation. Computation-related operations ({\tt{fusion, convolution, dynamic-update-slice}}) are mainly from kernel operations described in~\cref{sssec:kernel_op}. This fine-grained analysis shows that the observed superlinear behavior is due to the XLA compiler, which performs better at optimizing the TensorFlow data-structure when fewer operations are distributed to each core. This results in a more efficient computational graph with better fused operations that are MXU executable during run-time. We confirmed this through further numerical experiments with different partitions (along the $x$-$y$ plane), which does not reduce the number of operations per core but decrease the size of the operands in the operations. This resulted in sublinear speedup as the XLA compiler was not able to further optimize the operation on the fragmented data-structure of the computational graph.

For the second case, a superlinear scaling is retained up to 512 cores, and changes to a linear scaling for 2048 cores. This can be attributed to the fact that partitioning in the $x$- and $y$-directions is less efficient because the number of operations in the TensorFlow graph per core is no longer decreasing. The additional optimization from the XLA compiler when the number of operations is small no longer applies. 

\section{\label{SEC_CONCL}Conclusions} 
In this work, we have presented the development of a TensorFlow-based simulation framework for high performance scientific computing of fluid flows on TPU architectures. This simulation framework adopts a low-Mach number, variable-density formulation, which enables the simulation of a wide range of scientific and engineering flow problems. The simulation framework is implemented using TensorFlow's Python application programming interface and compiled into a TPU-executable graph using the domain-specific XLA compiler for acceleration. Aspects pertaining to the mapping of the data structure to the TPU hardware architecture were discussed and analyzed. To examine the impact of the machine round-off error of the algorithm, numerical analysis was performed and recommendations for spatial and temporal resolution were derived to minimize errors arising from the ML-specific floating-point arithmetic on TPUs. The simulation framework was applied to four test cases, which include 2D and 3D Taylor-Green vortex flow, decaying homogeneous isotropic turbulence, and a turbulent planar jet. Simulation results confirm that the solutions were not polluted by round-off errors and second-order accuracy was demonstrated. 

Assessment of scalability showed excellent weak scalability and superlinear strong scaling on up to 2048 TPU cores that constitute a full TPU pod. An analysis of code performed showed that this superlinear behavior is a consequence of compiler optimization of the data-structure resulting in a more efficient computational graph to improve the execution on matrix multiplier units during run time.

\section*{Acknowledgments}
We thank  Yusef Shafi, James Lottes, Damien Pierce, Tianjian Lu, Rasmus Munk Larsen, Sameer Agarwal, Blake Hechtman, Tao Wang, Anudhyan Boral, Carla Bromberg, and Zack Ontiveros at Google for valuable discussions and helpful comments.
\bibliographystyle{model1-num-names}
\bibliography{references}
\end{document}